\documentclass[pre,twocolumn,amssymb]{revtex4-1}
\usepackage{mathtools}
\usepackage{amsmath}
\usepackage{float}
\usepackage{graphicx,array}
\usepackage{color}
\usepackage{epsfig}
\usepackage{latexsym}
\usepackage{bm}
\usepackage{color}
\usepackage[normalem]{ulem}

\begin{document}

\newcommand{\rar}{\rightarrow}
\newcommand{\lar}{\leftarrow}
\newcommand{\rlh}{\rightleftharpoons}
\newcommand{\eref}[1]{Eq.~(\ref{#1})}%
\newcommand{\Eref}[1]{Equation~(\ref{#1})}%
\newcommand{\fref}[1]{Fig.~\ref{#1}} %
\newcommand{\Fref}[1]{Figure~\ref{#1}}%
\newcommand{\sref}[1]{Sec.~\ref{#1}}%
\newcommand{\Sref}[1]{Section~\ref{#1}}%
\newcommand{\aref}[1]{Appendix~\ref{#1}}%

\renewcommand{\ni}{{\noindent}}
\newcommand{\dprime}{{\prime\prime}}
\newcommand{\be}{\begin{equation}}
\newcommand{\ee}{\end{equation}}
\newcommand{\bea}{\begin{eqnarray}}
\newcommand{\eea}{\end{eqnarray}}
\newcommand{\nn}{\nonumber}
\newcommand{\bk}{{\bf k}}
\newcommand{\bQ}{{\bf Q}}
\newcommand{\q}{{\bf q}}
\newcommand{\s}{{\bf s}}
\newcommand{\bN}{{\bf \nabla}}
\newcommand{\bA}{{\bf A}}
\newcommand{\bE}{{\bf E}}
\newcommand{\bj}{{\bf j}}
\newcommand{\bJ}{{\bf J}}
\newcommand{\bs}{{\bf v}_s}
\newcommand{\bn}{{\bf v}_n}
\newcommand{\bv}{{\bf v}}
\newcommand{\la}{\left\langle}
\newcommand{\ra}{\right\rangle}
\newcommand{\dg}{\dagger}
\newcommand{\br}{{\bf{r}}}
\newcommand{\brp}{{\bf{r}^\prime}}
\newcommand{\bq}{{\bf{q}}}
\newcommand{\hx}{\hat{\bf x}}
\newcommand{\hy}{\hat{\bf y}}
\newcommand{\bS}{{\bf S}}
\newcommand{\cU}{{\cal U}}
\newcommand{\cD}{D}
\newcommand{\bR}{{\bf R}}
\newcommand{\pll}{\parallel}
\newcommand{\sumr}{\sum_{\vr}}
\newcommand{\cP}{{\cal P}}
\newcommand{\cQ}{{\cal Q}}
\newcommand{\cS}{{\cal S}}
\newcommand{\ua}{\uparrow}
\newcommand{\da}{\downarrow}
\newcommand{\red}{\textcolor {red}}
\newcommand{\black}{\textcolor {black}}
\newcommand{\1}{{\oldstylenums{1}}}
\newcommand{\2}{{\oldstylenums{2}}}
\newcommand{\mDelta}{\varepsilon}
\newcommand{\m}{\tilde m}
\def\lsim {\protect \raisebox{-0.75ex}[-1.5ex]{$\;\stackrel{<}{\sim}\;$}}
\def\gsim {\protect \raisebox{-0.75ex}[-1.5ex]{$\;\stackrel{>}{\sim}\;$}}
\def\lsimeq {\protect \raisebox{-0.75ex}[-1.5ex]{$\;\stackrel{<}{\simeq}\;$}}
\def\gsimeq {\protect \raisebox{-0.75ex}[-1.5ex]{$\;\stackrel{>}{\simeq}\;$}}

\title{\large{Time-dependent properties of run-and-tumble particles. II.: Current fluctuations }}

\author{Tanmoy Chakraborty}
\email{tanmoy.chakraborty@bose.res.in}  

\author{Punyabrata Pradhan}
\email{punyabrata.pradhan@bose.res.in}

\affiliation{Department of Physics of Complex Systems, S. N. Bose National Centre for Basic Sciences, Block-JD, Sector-III, Salt Lake, Kolkata 700106, India}

\begin{abstract}

     We investigate steady-state current fluctuations in two models of run-and-tumble particles (RTPs) on a ring of $L$ sites, for {\it arbitrary} tumbling rate $\gamma=\tau_p^{-1}$ and density $\rho$; model I consists of standard hardcore RTPs, while model II is an analytically tractable variant of model I, called long-ranged lattice gas (LLG). We show that, in the limit of $L$ large, the fluctuation  of cumulative current $Q_i(T,L)$ across $i$th bond in a time interval $T \gg 1/D$ grows first {\it subdiffusively} and then {\it diffusively} (linearly) with $T$, where $D$ is the bulk diffusion coefficient. Remarkably, regardless of the model details, the scaled bond-current fluctuations $D \langle Q_i^2(T,L) \rangle/2 \chi L \equiv {\cal W}(y)$ as a function of scaled variable $y=DT/L^2$ collapse onto a {\it universal} scaling curve ${\cal W}(y)$, where $\chi(\rho,\gamma)$ is the collective particle {\it mobility}. In the limit of small density and tumbling rate $\rho, \gamma \rightarrow 0$ with $\psi=\rho/\gamma$ fixed, there exists a scaling law: The scaled mobility $\gamma^{a} \chi(\rho, \gamma)/\chi^{(0)} \equiv {\cal H} (\psi)$ as a function of $\psi$ collapse onto a scaling curve ${\cal H}(\psi)$, where $a=1$ and $2$ in models I and II, respectively, and $\chi^{(0)}$ is the mobility in the limiting case of symmetric simple exclusion process (SSEP). For model II (LLG), we calculate exactly, within a   truncation scheme, both the scaling functions, ${\cal W}(y)$ and ${\cal H}(\psi)$. We also calculate spatial correlation functions for the current, and compare our theory with simulation results of model I; for both models, the correlation functions decay exponentially, with correlation length $\xi \sim \tau_p^{1/2}$ diverging with persistence time $\tau_p \gg 1$. Overall our theory is in excellent agreement with simulations and complements the findings of Ref. {\it arXiv:2209.11995}.  

\end{abstract}

\maketitle

\section{Introduction}

Characterizing the time-dependent properties of interacting self-propelled particles (SPPs), also known as active matter, has attracted a lot of attention in the past \cite{Ramaswamy2013, Bechinger}. SPPs convert chemical energy into directed or persistent motion (``run''); they move with a velocity $v$ and randomly change directions, or tumble, with a rate $\gamma$, thus breaking detailed balance at microscopic scales and driving the system out of equilibrium. Because of the delicate interplay between persistence and interactions, they display remarkable collective behavior, such as flocking \cite{Vicsek_1995}, clustering \cite{MIPS_2015}, ``giant" number fluctuation \cite{Narayanan_GNF}, and anomalous transport \cite{Takatori_2016,dolai2020}. Indeed, over the last couple of decades, significant effort has been made to better understand the emergent properties of active matter through studies of paradigmatic models, such as the celebrated Vicsek models \cite{Vicsek_1995}, active Brownian particles (ABPs) \cite{ABP_marchetti} and run-and-tumble particles (RTPs) \cite{RTP_tailleur}. However, despite numerous simulation and analytical studies in the past \cite{Ramaswamy2013, Bechinger}, theoretical characterization of the dynamic properties of these many-body systems  poses a major challenge and is still far from complete.

Understanding dynamic phenomena, such as density relaxation and current fluctuation, in out-of-equilibrium many-body systems is a fundamental problem in statistical physics. However, unlike the linear-response theory for equilibrium systems, a general theoretical formulation of the issue in a nonequilibrium setting is still lacking.
However, a deeper understanding of the problem for driven diffusive systems is gradually emerging through studies of macroscopic transport coefficients such as the collective- or {\it bulk-diffusion coefficient} and the {\it mobility}, respectively \cite{Derrida_PRL2004, MFT-RMP2015}. 
While the density relaxation is characterized through the bulk-diffusion coefficient, which is related to the relaxation rate of long-wavelength perturbations \cite{KLS-JSP1984}, the current fluctuation can be characterized through the collective particle mobility \cite{Derrida_PRL2004}.
In principle, the bulk-diffusion coefficient should be distinguished from the self-diffusion coefficient of a tagged particle \cite{Hanna_1981, DeMasi_JSP_2002, Tanmoy_condmat2022}; however the distinction, which can be quite striking especially in one dimension and also other models \cite{Mallick-PRE2014, Anirban-PRE2023}, is somewhat less emphasized in the context of active matter systems \cite{Speck-EPL2013, Pagonabarraga_PRL_2019}. One might expect the two transport coefficients to be connected \cite{Derrida-Gerschenfeld-JSP2009, Anirban-PRE2023}, but it is not clear how. It is worth noting here that the {\it cumulative} or the space-time integrated current across the entire system is nothing but the {\it cumulative} displacement of all (tagged) particles, in a given direction.
One of the primary objectives for our work is the theoretical characterization of the space-time integrated current fluctuations. Another goal is to understand the spatial correlations of (appropriately coarse-grained) current or, equivalently, ``velocity'', which has recently received significant attention in the contexts of coarsening in the Vicsek model \cite{Puri_EPJE_2020}, ordering dynamics in the ABPs \cite{Caprini_PRL_2020, Caprini_2020_PRR,Caprini_2021_SM,Szamel-EPL2021}, and other active matter systems \cite{Gov_PNAS2015, Cavagna_Nature_2017, Bertin_NatComm2020, Berthier_PRL2022}.

In the past decades, several analytical studies of current fluctuations for passive lattice gases have been conducted by using microscopic \cite{Derrida-Lebowitz-SSEP-ASEP, ZRP-Harris,Derrida-PRE2008, Derrida-Gerschenfeld-JSP2009, Derrida-Sadhu-JSTAT2016} as well as hydrodynamic frameworks \cite{Derrida_PRL2004, Bertini_PRL2005, Derrida-Gerschenfeld2-JSP2009}. However, their extension  to interacting SPPs is still a work in progress. In fact, unlike studies of tagged particle displacement fluctuations \cite{Suma-PRL2017, Banerjee-Cates-JSTAT2022, Benichou_PRL_2018, Benichou_PRL_2022}, there are few studies of current fluctuations in conventional models of SPPs, with the exception of an exact analysis for noninteracting RTPs \cite{Tirthankar_PRE_2020} and an approximate analysis for ABPs \cite{Limmer_ABP_2018}.  Recently, exact studies of fluctuations in ``weakly interacting'' RTPs \cite{Tailleur_PRL_2018}, which is governed by mean-field hydrodynamics, was carried out in Refs. \cite{Kafri-JSTAT2022, agranov_2023_scipost}. While weakly interacting RTPs undergo diffusion (symmetric hopping) with a {\it finite} rate, the run and tumble dynamics on the other hand occur with {vanishingly small} {\it system-size-dependent} rates, thus distinguishing the model from the {\it conventional or the standard} ones \cite{Evans-Blyhte-PRL2016, Soto_2014}. Indeed, in the latter models, the hopping dynamics is {\it independent} of system-size; as a result, spatial correlations are {\it finite}, and long-ranged for small tumbling rates. In such a case,  the mean-field description of Ref. \cite{Tailleur_PRL_2018, Kafri-JSTAT2022} would not be applicable  and a detailed microscopic study of the conventional, ``strongly interacting'', RTPs is desirable. 
It is worth noting that the order of  limits, tumbling rate $\gamma \rightarrow 0$ and systems size $L \rightarrow \infty$, is important and, depending on the order, there can be various distinct circumstances as described below. 
\\
\\
(1) As one takes the limit $\gamma \rightarrow 0$ first and then the limit $L \rightarrow \infty$ (by keeping drift velocity $v$ fixed), the system eventually goes into a state of {\it dynamical arrest} \cite{chandan_dasgupta_NatComm2020}, where dynamical activities cease and the system becomes frozen in time. 
\\
\\
(2) The tumbling rate $\gamma(L) \sim {\cal O}(L^{-\delta_1})$ and drift velocity $v(L) \sim {\cal O}(L^{-\delta_2})$ are taken to be system-size-dependent and one takes the limit $L \rightarrow \infty$, by keeping diffusion rate fixed; a specific case with $\delta_1 = 2$ and $\delta_2=1$ was considered in Refs. \cite{Tailleur_PRL_2018} and \cite{agranov_2023_scipost}. 
\\
\\
(3) On the other hand, in this work, we consider RTPs in the limit of $L$ large, with $\gamma$ fixed (diffusion rate is strictly zero) \cite{Tanmoy_condmat2022}. We are particularly interested in $\gamma \rightarrow 0$ limit, but, in that case,  the $L \rightarrow \infty$ limit is taken first; in other words, persistent length $l_p = v/ \gamma$ is finite, and large, but much smaller than the system size $L$, $1 \ll l_p \ll L$. We keep the drift velocity $v$ fixed.

In this paper, we use a microscopic approach to investigate spatio-temporal correlations of current in two models of strongly interacting run-and-tumble particles (RTPs): Model I - standard hardcore RTPs and model II - a hardcore long-ranged lattice gas, which is an analytically tractable idealized variant of model I. We study the models on a ring of $L$ sites for arbitrary tumbling rate $\gamma =\tau_p^{-1}$ and density $\rho$. Interestingly, despite having nontrivial many-body correlations, model II is amenable to ``first-principles'' analytical calculations, whereas model I is studied through Monte-Carlo simulations. 
We demonstrate that large-scale fluctuations in both models  can be characterized in terms of the two density- and tumbling-rate-dependent transport coefficients - the bulk-diffusion coefficient $D(\rho, \gamma)$ and the collective particle mobility $\chi(\rho, \gamma)$. Indeed, the current fluctuations and the mobility in model II are calculated exactly, within a previously introduced truncation scheme of Ref. \cite{Anirban-PRE2023}, and is expressed in terms of  distribution of gaps between two consecutive particles.
For convenience, we provide below a brief summary of our main findings.

\begin{itemize}

\item   {\it Spatial correlations of current.-- } We calculate steady-state spatial correlation function $C^{JJ}_{r} = \lim_{t \rightarrow \infty} \langle J_0(t) J_r(t) \rangle$ evaluated at two spatial points separated by distance $r$, with $J_i(t)$ being instantaneous current across bond $(i,i+1)$ at time $t$. The correlation functions decay exponentially, $C^{J J}_r \sim \exp(-r/\xi)$; the correlation length $\xi(\rho, \gamma)$ is  analytically calculated for LLG, with $\xi \sim \sqrt{\tau_p}$ diverging with persistence time $\tau_p$, thus providing a theoretical explanation of the findings in recent simulations and experiments \cite{Caprini_PRL_2020, Bertin_NatComm2020}.

\item \textit{Space-time integrated current fluctuations. -} We calculate fluctuations of space-time integrated current $Q_{tot} =\sum_{i=1}^L Q_i(T)$ or, equivalently, the cumulative displacement of all particles, during time interval $T$; here $Q_i(T) = \int_t^{t+T} dt J_i(t)$ is the cumulative bond current across $i$th bond $(i,i+1)$ during time $T$. We then study the collective {\it mobility} $\chi(\rho,\gamma) \equiv \lim_{L \rightarrow \infty} (1/2LT) \langle Q_{tot}^2\rangle$, which, in the limit of $ \rho, \gamma \rightarrow 0$, obeys a scaling law: The scaled mobility $\gamma^a \chi(\rho, \gamma)/\chi^{(0)}$ as a function of scaled variable $\psi=\rho/\gamma$ is expressed through a scaling function ${\cal H}(\psi)$; here $\chi^{(0)}=\rho (1-\rho)$ is the mobility in the limiting case of symmetric simple exclusion process (SSEP) \cite{Derrida_PRL2004}. The scaling function for LLG  is calculated analytically and, in the limit of strong persistence $\psi \gg 1$, it is shown to have an asymptotic behavior ${\cal H(\psi)} \sim \psi^{-3/2}$.

\item   \textit{Time-integrated bond-current fluctuations.--} Depending on the density- and tumbling-rate-dependent collective- or bulk-diffusion coefficient $D(\rho, \gamma)$ and system size $L$, we find three distinct time regimes for the fluctuations of time-integrated bond current $Q_i(T)$.
  (i) {\it Initial-time regime  $T \ll 1/D$:} The bond-current fluctuation $\langle Q_{i}^{2} \rangle(T,L)$ depends on the details of dynamical rules. It exhibits  linear (diffusive) growth for model II (LLG), whereas, for model I (standard RTPs), it crosses over from superdiffusive (even ballistic at low densities) to a diffusive growth as tumbling rate increases.
  (ii) {\it Intermediate-time regime $1/D \ll T \ll L^{2}/D$:}  The current fluctuation displays subdiffusive growth $\langle Q_{i}^{2} \rangle \sim \sqrt{T}$. 
  (iii)  {\it Long-time regime $L^{2}/D \ll T$:} The bond-current fluctuation $\langle Q_{i}^{2} \rangle \sim T$ grows diffuively (linear growth).
  So the qualitative behavior in regimes (ii) and (iii) is \textit{universal}, being independent of dynamical rules of the models; though the prefactors in the growth laws  are model-dependent.

\item \textit{Universal scaling of bond-current fluctuations.--} Remarkably, in the limit of $L, T \rightarrow \infty$ with scaled time $y=DT/L^{2}$ fixed and regardless of the dynamical rules of the models, the above mentioned behavior can be succinctly expressed through a {\it universal} scaling law for the scaled bond-current fluctuation $D(\rho, \gamma) \langle Q_{i}^{2} \rangle(T,L)/2 \chi L \equiv \mathcal{W}(y)$ as a function of $y=DT/L^2$ [see Eq.~\eqref{bond-current-fluc-3}]. For model II (LLG), the scaling function $\mathcal{W}(y)$ is calculated exactly within the truncation scheme.

\item \textit{Dynamic correlations of current -} We also calculate, numerically for model I  (standard RTPs) and analytically for model II (LLG), the two-point dynamic correlation function $\mathcal{C}^{J J}_{0}(t) = \langle J_i(0) J_i(t) \rangle$ for instantaneous current $J_i(t)$. For model II, by using our microscopic dynamical calculations, we derive the dynamic correlation function $\mathcal{C}^{J J }_{0}(t) \sim -t^{-3/2}$, which is shown to have a long-time power-law tail, with the correlations actually being {\it negative}. 


\end{itemize}

The paper is organized as follows: In Sec.~\ref{Sec:model}, we introduce two models of hardcore RTPs. In sec.~\ref{Sec:current_decomposition} we describe the formal procedure for  decomposition of current into ``slow" (diffusive) and ``fast" (noise-like) components. Then in Sec.~\ref{Sec:truncation}, we introduce a truncation scheme, which allows us to calculate the spatio-temporal correlations of time-integrated currents. In Secs.~\ref{Sec:space-time_inst}, \ref{Sec:space-time_fluc}, we investigate spatio-temporal correlations of instantaneous and fluctuating currents, respectively. Next in Sec.~\ref{Sec:fluc_total}, we characterize fluctuations of total current, leading to the characterization of the collective particle mobility $\chi(\rho,\gamma)$; here we also find a scaling law in the limit of strong persistence and dilute regime. In Sec~\ref{Sec:bond_current_fluc}, we characterize bond-current fluctuation and find another scaling law, presumably universal, in the large-scale diffusive limit. Finally, we summarize the paper in sec.~\ref{Sec:conclusion} with some concluding remarks.

\section{Model Description}\label{Sec:model}

We consider two minimal models of interacting RTPs on a one-dimensional periodic lattice of $L$ sites where the number of particles $N$ is conserved with density $\rho=N/L$. In both models, particles obey hardcore constraint, i.e., a site can be occupied by at most one particle and also the crossing between particles is not allowed. We denote the occupation variable $\eta_i=1$ or $0$, depending on whether the site is occupied or not, respectively.

\begin{figure}[tpb]
           \centering
         \includegraphics[width=0.75\linewidth]{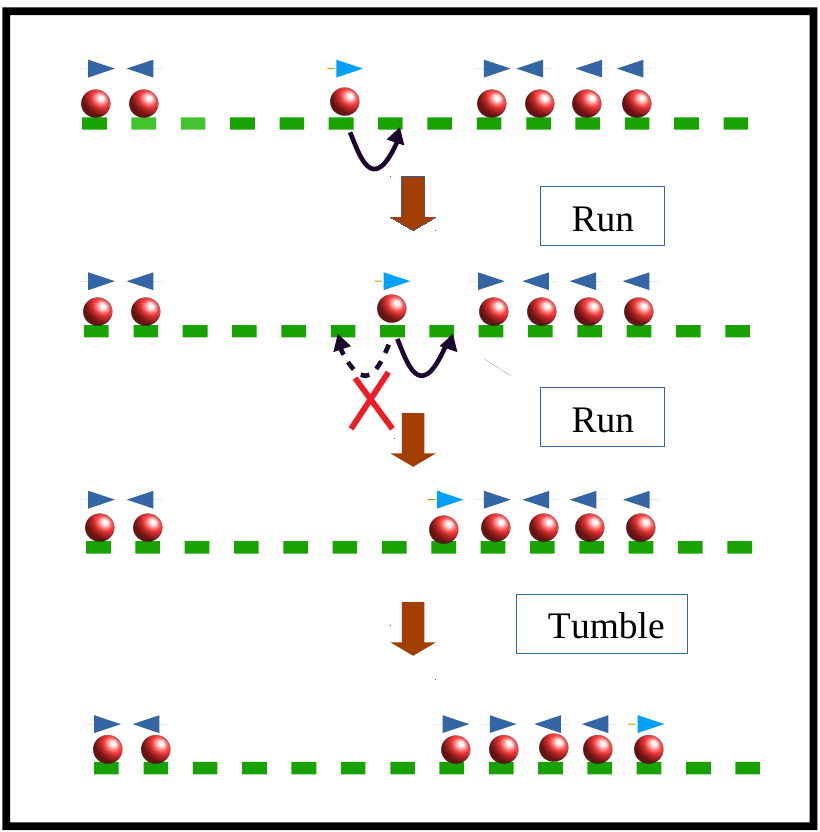}\\
         \includegraphics[width=0.75\linewidth]{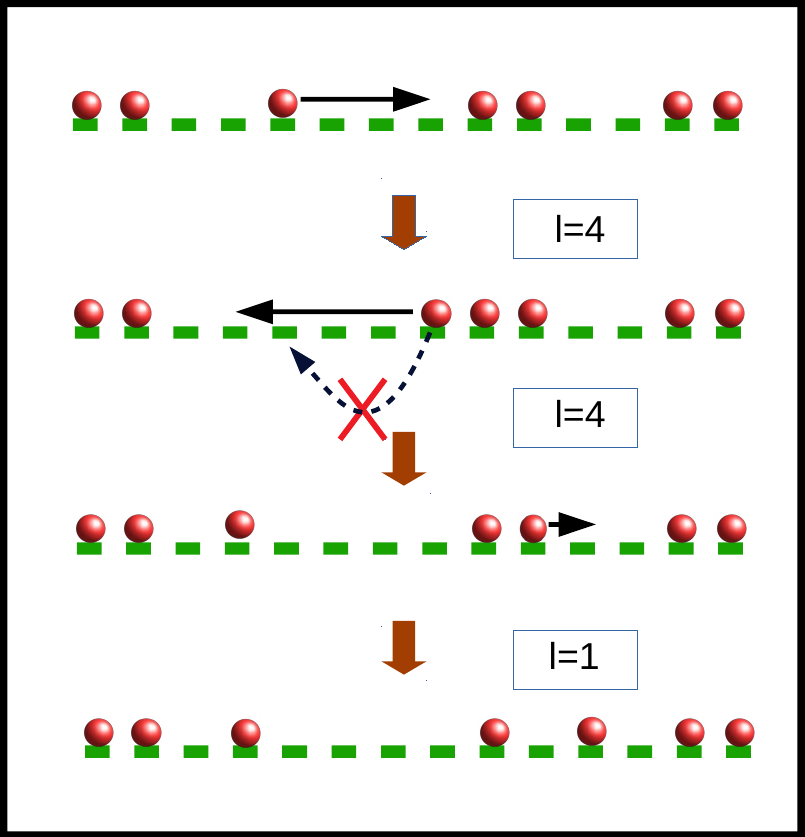}
         \caption{\textit{Schematic diagram of model I and II-} The top panel illustrates the typical dynamics of model I, which is composed of standard hardcore RTPs (red circles) on a one-dimensional lattice (green rectangles). RTPs move along the associated spin indicated by the arrows above them. In the bottom panel, we demonstrate the dynamics of hardcore particles (red circle) in model II (i.e., LLG) on one dimensional lattice (green rectangle). Particles hop symmetrically by length $l$ drawn from distribution $\phi(l) \sim e^{-l/l_p}$. The ``crosses" in both panels indicate the impossibility of the time-reversed moves, as shown by the dotted arrows, and thus the violation of detailed balance at the microscopic level in both models I and II.}
          \label{fig:model_diagram}
 \end{figure}

\subsection*{Model I: Standard hardcore RTPs}

We consider standard hardcore RTPs (see the schematic diagram in the \textit{top-panel} of Fig.~\ref{fig:model_diagram}) introduced in Ref.~\cite{Soto_2014}. In this model, in addition to the occupation variable, a spin $s=\pm 1$ is assigned to each particle, with $s=1$ and $s=-1$ represent its rightward and leftward orientations, respectively. The continuous-time stochastic dynamics is specified below. 
  \\\\
  (A) \textit{Run:} With unit rate, a particle hops, along its spin direction, to its nearest neighbor provided that the destination site is vacant.
  \\\\
  (B) \textit{Tumble:} With rate $\gamma = \tau_p^{-1}$, a particle changes its spin orientation, $s \rightarrow -s$.
  \\\\
Clearly particles retain their spin orientation over a time scale of  the persistence time $\tau_p$ and, during this time, they exhibit ballistic motion with a constant speed $v$ [note that $v=1$, set by rule (A)] along the vacant stretch available in the direction of its spin.

\subsection*{Model II: Hardcore long-ranged lattice gas (LLG)}

Because of the additional spin variable,  model I proves to be difficult to deal with analytically. To address the issue, we also explore  a simpler idealized variant of hardcore RTPs, called long-ranged lattice gas (LLG) \cite{tanmoy_2020, Tanmoy_condmat2022}, which is amenable to analytical studies. The long-range hopping mimics ballistic motion (``run'') of individual RTPs, having a characteristic run-length - the {\it persistence length} $l_p=v/\gamma$. Indeed, model II (LLG) is motivated by the fact that, on the time scale of $\tau_p$, a single RTP would hop, on average, by the typical length $l_p$. 

The precise dynamical rules of model II (see the schematic diagram in the \textit{bottom-panel} of Fig.~\ref{fig:model_diagram}) are as following. With unit rate, a particle attempts to hop symmetrically by length $l$, drawn from a distribution $\phi(l)$. The hop is successful provided that, along the hopping direction, there is a vacant stretch, called {\it gap}, of size $g$ which is of at least size $l$ (i.e., $g \geq l$); otherwise (i.e., for $g < l$), due to the hardcore constraint, the particle traverses the entire stretch and sits adjacent to its nearest occupied site. The hop-length distribution $\phi(l)$ can be arbitrary. However, in order to compare the two models I and II, we chose $\phi(l)$ to be an exponential function,
\begin{eqnarray}\label{hop-length}
\phi(l) = A \exp(-l/l_p),
\end{eqnarray}
where the normalization constant $A=(1-e^{-1/l_p})$ as hop-lengths $l=0,1, 2, \dots $ are discrete.

\section{Theory: Model II }

In this section, we develop a microscopic theoretical framework to analytically calculate current fluctuations in model II (hardcore LLG). We also compare our analytic results with that obtained from direct Monte Carlo simulations of both models I and II.

\subsection{ Decomposition of current: Slow and fast components } 

\label{Sec:current_decomposition}
 
To begin with, we first define cumulative (time-integrated) bond current $Q_{i}(T)$, which is total current across a bond $(i, i+1)$ in a time interval $T$. On the other hand, instantaneous current $J_{i}(t)$ is defined as
\begin{eqnarray}\label{inst_and_integrated_current}
J_{i}(t) \equiv  \lim_{\Delta t \rightarrow 0} \frac{\Delta Q_i }{\Delta t},
\end{eqnarray}
where $\Delta Q_i = \int_t^{t + \Delta t} dt J_i(t) $ is cumulative bond current in time interval $\Delta t$.
Note that, while we investigate fluctuation properties of both the quantities $Q_i(t)$ and $J_i(t)$,  in simulations it is statistically more efficient to calculate the averages related to the time-integrated current $Q_i(t)$ than the instantaneous one $J_i(t)$.

However, before  discussing second moment (fluctuations) of currents, in this section we first investigate their average behavior and set the notations, required in the subsequent part of the paper. To this end, let us  define the following stochastic variables,
  \begin{eqnarray}
 \mathcal{U}_{i+l}^{(l)} &\equiv& \overline{\eta}_{i+1} \overline{\eta}_{i+2} \dots \overline{\eta}_{i+l} , \\ 
 \mathcal{V}_{i+l+1}^{(l+2)} &\equiv& \eta_{i}\overline{\eta}_{i+1} \overline{\eta}_{i+2} \dots \overline{\eta}_{i+l}\eta_{i+l+1} , 
 \end{eqnarray}
 where $\bar{\eta}_i=(1-\eta_i)$,  $\mathcal{U}^{(l)}$ and $\mathcal{V}^{(l+2)}$ are indicator functions for a single site being vacant, $l$ consecutive sites being vacant and a vacancy cluster to be of size $l$, respectively. Note that, whenever a particle performs long-range hop of length $l$, it contributes a unit current at each bond in the stretch between the departure and destination sites; that is, for a rightward (leftward) hop, current across {\it all} bonds in that stretch increases (decreases) by unity. By considering this, the continuous-time evolution for time-integrated current $Q_i(t)$ in an infinitesimal time interval $[t, t+dt]$ can be written as

\begin{eqnarray} 
 Q_i(t+dt) = 
\left\{
\begin{array}{ll}
\vspace{0.15 cm}
 Q_i(t)  + 1,            ~~~  & {\rm prob.}~~~ \mathcal{P}^{R}_{i}(t) dt , \\
\vspace{0.15 cm}
 Q_i(t) - 1,            ~~~  & {\rm prob.}~~~ \mathcal{P}^{L}_{i}(t) dt, \\
 \vspace{0.15 cm}
 Q_i(t),                ~~~  & {\rm prob.}~~~~  1 - (\mathcal{P}^{R}_{i} + \mathcal{P}^{L}_{i}) dt, \\
\end{array}
\right.
\label{Q_update_eq}
\end{eqnarray}
where $\mathcal{P}^{R}_{i} dt$ and $\mathcal{P}^{L}_{i} dt$ are probabilities of the hopping events and the corresponding rates are given by \cite{Tanmoy_condmat2022} 
\begin{eqnarray}
  \label{pr}
  \mathcal{P}^{R}_{i} &\equiv& \frac{1}{2}\sum_{l=1}^{\infty} \phi(l) \left[\sum_{k=1}^{l} \left(\mathcal{U}_{i+k}^{(l)} - \mathcal{U}_{i+k}^{(l+1)}\right) + \sum_{g=1}^{l-1}\sum_{k=1}^{g}\mathcal{V}_{i+k+1}^{(g+2)}\right].\nonumber \\ \\
\mathcal{P}^{L}_{i} &\equiv & \frac{1}{2}\sum_{l=1}^{\infty} \phi(l) \left[\sum_{k=1}^{l} \left(\mathcal{U}_{i+k-1}^{(l)} - \mathcal{U}_{i+k}^{(l+1)}\right) + \sum_{g=1}^{l-1}\sum_{k=1}^{g}\mathcal{V}_{i+k}^{(g+2)}\right]. \nonumber \\  \label{pl}
\end{eqnarray}
By using the above microscopic update rules and doing some straightforward algebraic manipulations, the average instantaneous current $\left \langle J_{i}(t) \right \rangle$ can be written as \cite{Tanmoy_condmat2022}
\begin{eqnarray}
  \label{time-derivative-int-current_2}
\left \langle J_{i}(t) \right \rangle &=& \frac{1}{2}\sum_{l=1}^{\infty}\phi(l)\left[\sum_{g=1}^{l-1}\Big( \langle \mathcal{V}_{i+g+1}^{(g+2)} \rangle - \langle \mathcal{V}_{i+1}^{(g+2)} \rangle \right) \nonumber \\ && \hspace{2.75 cm} + \left( \langle \mathcal{U}_{i+l}^{(l)} \rangle -\langle \mathcal{U}_i^{(l)}\rangle \Big)\right].
\end{eqnarray}
Note that, in the above equation, the average current $\langle J_{i}(t) \rangle$ is written as a (generalized) gradient of the observables $\langle \mathcal{V}^{(g+2)}\rangle (\rho)$ and $\langle\mathcal{U}^{(l)} \rangle (\rho)$, both of which depend on (local) density and, of course, tumbling rate. By making use of the Taylor's expansion, we can write the average current explicitly in terms of (discrete) gradient of (local) density  \cite{Tanmoy_condmat2022},
\begin{eqnarray}
  \label{cont-eqn}
\left \langle J_i(t) \right \rangle \simeq - D(\rho, \gamma) [\langle \eta_{i+1}(t) \rangle - \langle \eta_i(t) \rangle]  
\end{eqnarray}
where $\rho_i(t) = \langle \eta_i(t) \rangle$ is the local density and the bulk-diffusion coefficient $D(\rho, \gamma)$ \cite{Tanmoy_condmat2022}
\begin{eqnarray}
  \label{bulk-diffusivity-LH_sm}
 D(\rho,\gamma)&=& -\frac{1}{2}\sum_{l=1}^{\infty} \phi(l) \frac{\partial}{\partial \rho} \left[\sum_{g=1}^{l-1} g \langle \mathcal{V}^{(g+2)} \rangle(\rho) + l \langle \mathcal{U}^{(l)} \rangle (\rho) \right], \nonumber \\
 \end{eqnarray}
is a function of the global density $\rho$ and tumbling rate $\gamma$.

As the system is homogeneous in the steady-state, the gradients in Eq.~\eqref{time-derivative-int-current_2} simply vanish, implying that the system has {\it zero} steady-state current. However, on the level of fluctuations, the (stochastic) instantaneous current is in fact {\it nonzero} even in the steady-state. Now, to characterize fluctuations appropriately \cite{Derrida_PRL2004, MFT-RMP2015}, we decompose the total instantaneous current  into two components: A (hydrodynamic) diffusive current $J^{(D)}_{i}$, which, though stochastic, relaxes very slowly, and a fluctuating or noise current $J^{(fl)}_{i}$, which relaxes very fast. In other words, we write the instantaneous current as the sum of these slow and fast components,
\begin{eqnarray}
\label{current_decompose}
J_{i}(t) = J^{(D)}_{i}(t) + J^{(fl)}_{i}(t),
\end{eqnarray}
where we identify, by using Eq. \eqref{time-derivative-int-current_2}, the  diffusive current
\begin{eqnarray}\label{diffusive_current}
J^{(D)}_i &\equiv& \frac{1}{2} \sum_{l=1}^{\infty}\phi(l) \Bigg[\sum_{g=1}^{l-1}\left( \mathcal{V}_{i+g+1}^{(g+2)} - \mathcal{V}_{i+1}^{(g+2)} \right) \nonumber \\ &&  \hspace{2.5 cm}+ \left( \mathcal{U}_{i+l}^{(l)} -\mathcal{U}_i^{(l)} \right)\Bigg].
\end{eqnarray}
Indeed, as we derive later [see Eqs. \eqref{inst_current_temp_correlation_asymptotic} and \eqref{fluc_current_correlation}], time-dependent correlation function for the diffusive current $J^{(D)}_i$ (also the total current $J_i$) has a power-law tail, whereas that for the fluctuating (noise) current $J^{(fl)}_i$ is delta-correlated. 
Also, comparing Eqs.~\eqref{time-derivative-int-current_2}, \eqref{current_decompose} and \eqref{diffusive_current}, we find that average fluctuating current is simply zero, 
\begin{eqnarray}
\langle J^{fl}_{i}(t) \rangle =0.
\end{eqnarray}
However the space-time correlations of $J_i^{fl}$ has nontrivial spatial structures [see Eq.~\eqref{fluc_current_correlation}] and, in the subsequent section, they are analytically calculated  by using a truncation scheme, which we discuss next.

\subsection{Spatio-temporal correlations of current}\label{Sec:truncation}

We consider the time-integrated currents $Q_{r}(t')$ and $Q_{0}(t)$, which are measured up to times $t'$ and $t$ $(t' > t)$ across bonds $(r, r+1)$ and $(0, 1)$, respectively, where the bonds are spatially separated by a distance $r$. In this section, we investigate the space-time-dependent correlation function for bond current $Q_i(t)$,
\begin{eqnarray}
\label{current_correlation_defn}
\mathcal{C}^{QQ}_{r}(t',t) &=& \left\langle Q_{r}(t') Q_{0}(t) \right\rangle_{c}, \nonumber \\ &=& \left\langle Q_{r}(t') Q_{0}(t) \right\rangle - \left\langle Q_{r}(t')  \right\rangle \left\langle  Q_{0}(t) \right\rangle.
\end{eqnarray}
As we choose $t' > t$, it is easy to see that, in an infintesimal time-interval $[t', t'+dt']$,  $Q_{0}(t)$ remains constant and any change in $\mathcal{C}^{QQ}_{r}(t',t)$ occurs solely  due to the change in $Q_{r}(t')$. Now, by using infinitesimal update rules Eq. \eqref{Q_update_eq}, we write down the time-evolution equation for $\mathcal{C}^{QQ}_{r}(t',t)$ as given below (see Appendix~A for details), 
\begin{eqnarray}
  \label{time-evolution-Q-Q-general}
  \frac{d}{dt'}\mathcal{C}^{QQ}_{r}(t',t)  &=& \frac{1}{2}\sum_{l=1}^{\infty}\phi(l)\Big[ \Big\{ \mathcal{C}^{\mathcal{U}^{(l)}Q}_{r+l}(t',t) - \mathcal{C}^{\mathcal{U}^{(l)}Q}_{r}(t',t) \Big\} \nonumber \\ && \hspace{-0.75 cm} + \sum_{g=1}^{l-1} \Big\{\mathcal{C}^{\mathcal{V}^{(g+2)}Q}_{r+g+1}(t',t) - \mathcal{C}^{\mathcal{V}^{(g+2)}Q}_{r+1}(t',t) \Big\}\Big].
\end{eqnarray}
In other words, we have the following identity,
\begin{eqnarray}
\frac{d}{dt'}\mathcal{C}^{QQ}_{r}(t',t)  = \left \langle J^{(D)}_{r}(t') Q_{0}(t)\right \rangle_{c},
\label{time-evolution-Q-Q-general1}
\end{eqnarray}
where $J^{(D)}_{r}$ is the diffusive current at $r$th bond and time $t$.
Note that Eq.~\eqref{time-evolution-Q-Q-general} for two-point current correlation is exact  and has been expressed as the gradients of two nontrivial multi-point correlation functions,
\begin{eqnarray}
\mathcal{C}^{\mathcal{U}^{(l)}Q}_r(t',t)=\langle \mathcal{U}^{(l)}_r(t') Q_{0}(t) \rangle_{c}, \\
\mathcal{C}^{\mathcal{V}^{(g+2)}Q}_r(t',t)=\langle \mathcal{V}^{(g+2)}_{r}(t') Q_0(t) \rangle_{c}.
\end{eqnarray}
However, the difficulty arises here because the two-point correlation in eq. \eqref{time-evolution-Q-Q-general} actually involves various multi-point correlation functions, which must be now calculated in order to determine $\mathcal{C}^{QQ}_{r}(t',t)$. Not surprisingly, the hierarchy involving time-evolution of $\mathcal{C}^{\mathcal{U}^{(l)}Q}_{r}(t',t)$ and $\mathcal{C}^{\mathcal{V}^{(g+2)}Q}_{r}(t',t)$ does not close, making exact calculations extremely difficult.

To address the above mentioned difficulty, in this paper we  propose a truncation  scheme that, though approximate, allows us to close the above hierarchy and write the time-evolution of the two-point current correlations in terms of two-point correlations involving only current and density, which, interestingly, close onto themselves.
Indeed, when the fluctuations of local density around the steady state are small, on a long time scale the variables $\mathcal{V}^{(g+2)} $ and $ \mathcal{U}^{(l)} $ at a particular time, appearing in Eq.~\eqref{diffusive_current}, are ``slave'' to the local density and, as a result, the diffusive current could be approximately written in the form of a ``microscopic'' version of the Fick's law \cite{Derrida-Sadhu-JSTAT2016}, which is evident from Eq. \eqref{cont-eqn},
\begin{eqnarray}
  \label{closure_approximation}
J^{(D)}_{r}(t') \simeq D(\rho, \gamma) [ \eta_{r}(t') - \eta_{r+1}(t') ],
\end{eqnarray}
where we have simply used $D[\rho_{r}(t),\gamma] \simeq D(\rho,\gamma)$; the symbol ``$\simeq$'' in Eq. \eqref{closure_approximation} should rather be interpreted as an ``equivalence'', not an ``equality'', between the random variables there, unless one takes explicit averages. The precise implication of the above equivalence relation in Eq. \eqref{closure_approximation}, which has been used in the subsequent calculations, is the following.
We can simply write the correlation function for diffusive current $J^{(D)}_{r}(t')$ and any other stochastic variable $B(t)$ in terms of correlations between local density and the variable $B$,
\begin{eqnarray}
\left\langle J^{(D)}_{r}(t') B(t) \right \rangle_{c} \hspace{-0.0 cm} \simeq -D(\rho, \gamma) \Delta_{r}\left\langle \eta_{r}(t') B(t)\right \rangle_{c},
\label{TS}
\end{eqnarray}
where $\Delta_{r} h_r = h_{r+1} - h_r$ is the forward difference operator. Following the above truncation scheme Eq. \eqref{TS}, the time evolution of the current correlations in Eq.~\eqref{time-evolution-Q-Q-general1} greatly simplifies as the gradient of two-point density-current correlation function $\mathcal{C}^{\eta Q}_{r}(t',t)=\langle \eta_{r}(t') Q_{0}(t) \rangle_{c}$, which, as shown below, immediately closes the hierarchy. We thus rewrite Eq.~\eqref{time-evolution-Q-Q-general1} as
\begin{eqnarray}
\label{time-evolution-Q-Q-general_2}
\frac{d}{dt'}\mathcal{C}^{QQ}_{r}(t',t)  \simeq -D(\rho, \gamma) \Delta_{r} \mathcal{C}^{\eta Q}_{r}(t',t),
\end{eqnarray}
whereas the time-evolution of the density-current correlation $\mathcal{C}^{\eta Q}_{r}(t',t)$ can be written as [for detials see  Appendix~B],
\begin{eqnarray}\label{time-evolution-eta-Q-diff_time}
\frac{d}{dt'}\mathcal{C}^{\eta Q}_{r}(t',t)  = D(\rho, \gamma) \Delta^{2}_{r} \mathcal{C}^{\eta Q}_{r}(t',t).
\end{eqnarray}
Interestingly, the  SSEP, despite having a product-measure steady state and the bulk-diffusion coefficient being independent of density \cite{Derrida-Sadhu-JSTAT2016}, shares a similar structure of the density and current correlations for LLG, which have nonzero spatial correlations, albeit. In order to further simplify the time-evolution equations governing the two-point correlations, we represent the correlation functions in the Fourier space by using the transformation,
\begin{eqnarray}\label{Fourier_transform}
\tilde{\mathcal{C}}_{q}^{AB}(t',t)=\sum_{r=0}^{L-1}\mathcal{C}_{r}^{AB}(t',t) e^{iqr}.
\end{eqnarray} 
The inverse Fourier transform  is given by
\begin{eqnarray}\label{Inverse_Fourier_transform}
\mathcal{C}_{r}^{AB}(t',t)=\frac{1}{L}\sum_{q}\tilde{\mathcal{C}}_{q}^{AB}(t',t) e^{-iqr},
\end{eqnarray}
where
\begin{eqnarray}
q=\frac{2\pi n}{L},
\end{eqnarray}
and $n=0, 1, 2, \dots, (L-1)$. We rewrite Eqs.~\eqref{time-evolution-Q-Q-general_2} and \eqref{time-evolution-eta-Q-diff_time} in terms of the time evolution of the respective Fourier modes,
\begin{eqnarray}\label{time-evolution-Q-Q-diff-time-fourier}
\frac{d}{dt'}\mathcal{\tilde{C}}^{QQ}_{q}(t',t)  = D(\rho, \gamma) \left(1-e^{-i q} \right) \mathcal{\tilde{C}}^{QQ}_{q}(t',t),
\end{eqnarray}
and
\begin{eqnarray}
\label{time-evolution-eta-Q-diff-time-fourier}
\frac{d}{dt'}\mathcal{\tilde{C}}^{\eta Q}_{q}(t',t)  = -D(\rho, \gamma) \lambda_{q} \mathcal{\tilde{C}}^{\eta Q}_{q}(t',t),
\end{eqnarray}
where $\lambda_{q}$ is given by
\begin{eqnarray}\label{eigen-value}
\lambda_{q}=2\left(1- \cos q \right). 
\end{eqnarray}
By integrating Eqs.~\eqref{time-evolution-Q-Q-diff-time-fourier} and \eqref{time-evolution-eta-Q-diff-time-fourier}, we express the unequal-time correlation functions in the following forms,
\begin{eqnarray}\label{Q-Q_solution1}
\mathcal{\tilde{C}}^{QQ}_{q}(t',t) &=& D(\rho,\gamma) \hspace{-0.1 cm} \int_{t}^{t'} \hspace{-0.25 cm} dt'' \left(1-e^{-i q} \right) \mathcal{\tilde{C}}^{\eta Q}_{q}(t'',t) + \mathcal{\tilde{C}}^{QQ}_{q}(t,t), \nonumber \\ \\
\label{eta-Q_solution1}
\mathcal{\tilde{C}}^{\eta Q}_{q}(t'',t) &=& e^{-\lambda_q D(\rho,\gamma) (t''-t)} \mathcal{\tilde{C}}^{\eta Q}_{q}(t,t), 
\end{eqnarray}
where $t'' \geq t$. It is now clear  that, in order to evaluate the unequal-time correlation functions, $\mathcal{C}^{QQ}_{r}(t',t)$ and $\mathcal{C}^{\eta Q}_{r}(t',t)$, one must first calculate their equal-time counterparts, which we  do next.

\subsubsection{Equal-time density-current correlation $\mathcal{\tilde{C}}^{\eta Q}_{q}(t,t)$}

We have already seen in Eq.~\eqref{Q-Q_solution1} that the density-current correlation function $\mathcal{\tilde{C}}^{\eta Q}_{q}$ plays an important role in determining $\mathcal{\tilde{C}}^{QQ}_{q}$. We therefore proceed with the calculation of the equal-time density-current correlation function $\mathcal{\tilde{C}}^{\eta Q}_{q}(t,t)$; for details of the following calculations, see Appendix~C. 
Starting from microscopic dynamical rules, we obtain the time-evolution of $\mathcal{C}^{\eta Q}_{r}(t,t)$, which, in terms of the Fourier modes, satisfies the following equation,
\begin{eqnarray}\label{eta1_Q2_evolution_compact}
\left(\frac{d}{dt} + D(\rho,\gamma) \lambda_{q} \right)\mathcal{\tilde{C}}^{\eta Q}_{q}(t,t)=\mathcal{\tilde{S}}^{\eta Q}_{q}(t).
\end{eqnarray}
Here the  source term $\mathcal{\tilde{S}}^{\eta Q}_{q}(t)$  is given by
\begin{eqnarray}\label{source-eta-Q}
\mathcal{\tilde{S}}^{\eta Q}_{q}(t) &=& \frac{1}{(1-e^{-iq})}\left[D(\rho,\gamma) \lambda_{q} \mathcal{\tilde{C}}^{\eta \eta}_{q}(t,t) -f_{q}(t)\right], 
\end{eqnarray}
where $f_{q}(t)$ is directly related to the gap-distribution  $P(g, t)$ of the system and is given by
\begin{eqnarray}\label{f_{q}}
f_{q}(t)=\rho \sum_{l=1}^{\infty}\phi(l)\left[\sum_{g=1}^{l-1}  \lambda_{gq}P(g, t) +  \lambda_{lq}\sum_{g=l}^{\infty}P(g, t)\right].
\end{eqnarray}
We can now solve for $\mathcal{\tilde{C}}^{\eta Q}_{q}(t,t)$ by integrating Eq.~\eqref{eta1_Q2_evolution_compact} and the  solution is given by
\begin{eqnarray}\label{eta1-Q2_sametime}
\mathcal{\tilde{C}}^{\eta Q}_{q}(t,t)=\int_{0}^{t}dt''' e^{-\lambda_q D(\rho,\gamma)(t-t''')}\mathcal{\tilde{S}}^{\eta Q}_{q}(t'''),
\end{eqnarray}
which, upon substitution in Eq.~\eqref{eta-Q_solution1}, leads to the unequal-time density-current correlation function,
\begin{eqnarray}\label{eta1-Q2_diff_time}
\mathcal{\tilde{C}}^{\eta Q}_{q}(t'',t)=\int_{0}^{t}dt''' e^{-\lambda_q D(\rho,\gamma)(t''-t''')}\mathcal{\tilde{S}}^{\eta Q}_{q}(t''').
\end{eqnarray}
Note that, in the above analysis, $\mathcal{\tilde{S}}^{\eta Q}_{q}(t)$, hence $\mathcal{\tilde{C}}^{\eta Q}_{q}(t,t)$ and $\mathcal{\tilde{C}}^{\eta Q}_{q}(t'',t)$, have been expressed in terms of the equal-time density-density correlation function $\mathcal{\tilde{C}}^{\eta\eta}_{q}(t,t)$, which remains to be the only unknown quantity so far and has to be determined. To this end, we first derive the time-evolution equation for the correlation function $\mathcal{C}^{\eta \eta}_{r}(t,t)= \langle \eta_r(t) \eta_0(t) \rangle_{c}$, in the real space; for details,  see Appendix D. Using Fourier transform of Eq.~\eqref{Fourier_transform}, we find the time-evolution equation for  Fourier modes $\mathcal{\tilde{C}}^{\eta\eta}_{q}(t,t)$,
\begin{eqnarray}\label{eta1_eta2_evolution_compact}
\left(\frac{d}{dt} + 2 D(\rho,\gamma) \lambda_{q} \right) \mathcal{\tilde{C}}^{\eta \eta}_{q}(t,t) = \mathcal{\tilde{S}}^{\eta \eta}_{q}(t),
\end{eqnarray} 
where the  source term $\mathcal{\tilde{S}}^{\eta \eta}_{q}=f_{q}$.
One can now solve Eq.~\eqref{eta1_eta2_evolution_compact} to obtain the time-dependent solution of $\mathcal{\tilde{C}}^{\eta \eta}_{q}(t,t)$. Since we want the dynamic density-density correlation function to be evaluated in the steady state, we simply drop its time dependence and set $d \mathcal{\tilde{C}}^{\eta \eta}_{q}(t,t)/ dt=0 $; consequently, we have from Eq.~\eqref{eta1_eta2_evolution_compact},
\begin{eqnarray}\label{static-density-correlation}
2D(\rho,\gamma)\lambda_q \mathcal{\tilde{C}}^{\eta \eta}_{q}=\mathcal{\tilde{S}}^{\eta \eta}_{q}=f_{q}.
\vspace{0.25 cm}
\end{eqnarray}
The above Eq.~\eqref{static-density-correlation} provides the solution for the static density-density correlation function and $f_q$ is then obtained by replacing $P(g,t)$ by its steady-state value $P(g)$ in Eq.~\eqref{f_{q}}. Upon substituting the static $\mathcal{\tilde{C}}^{\eta \eta}_{q}$ in Eq.~\eqref{source-eta-Q}, the source term  $\mathcal{\tilde{S}}^{\eta Q}_{q}$ also becomes time-independent and thus the solution is given by
\begin{eqnarray}\label{eta_q_source_simplified}
\mathcal{\tilde{S}}^{\eta Q}_{q}&=&-\frac{f_{q}}{2(1-e^{-iq})}.
\end{eqnarray}
Using this particular form of $\mathcal{\tilde{S}}^{\eta Q}_{q}$ in Eq.~\eqref{eta1-Q2_diff_time}, we finally obtain the equal- as well as unequal-time density-current correlation function $\mathcal{\tilde{C}}^{\eta Q}_{q}$ in the steady state, 
\begin{eqnarray} 
\mathcal{\tilde{C}}^{\eta Q}_{q}(t,t)=\frac{-f_{q}}{2D(\rho,\gamma) \lambda_q (1-e^{-iq})}\left(1 - e^{-\lambda_q D(\rho,\gamma)t}\right),\label{density_current_equal_soln} \\
\mathcal{\tilde{C}}^{\eta Q}_{q}(t'',t)=\frac{-f_{q}e^{-\lambda_q D(\rho,\gamma)t''}}{2D(\rho,\gamma)\lambda_q(1-e^{-iq})}\left(e^{\lambda_q D(\rho,\gamma)t} - 1\right),\label{density_current_unequal_soln}
\end{eqnarray}
where $t'' \geq t$. It should be noted that, by substituting Eq. \eqref{density_current_unequal_soln} in Eq.\eqref{Q-Q_solution1}, one can readily obtain the first term of the unequal-time current-current correlation function $\mathcal{\tilde{C}}^{Q Q}_{q}(t'',t)$. In the next section, we focus on another equal-time correlation function $\mathcal{C}^{Q Q}_{r}(t,t)$,  which is the  quantity to be considered while calculating the two-point space-time correlation function $\mathcal{C}^{Q Q}_{r}(t',t)$.

\subsubsection{Equal-time current-current correlation $\mathcal{C}^{QQ}_{r}(t,t)$}

To calculate equal-time current-current correlation $\mathcal{C}^{Q Q}_{r}(t,t)$, we first derive its time-evolution equation, which, upon applying the closure scheme as in Eq.~\eqref{closure_approximation}, leads to the solution for $\mathcal{C}^{QQ}_{r}(t,t)$ having the following closed form expression,
\begin{eqnarray}\label{Q1_Q2_evolution_compact}
\mathcal{C}^{QQ}_{r}(t,t) &=& \frac{D}{L} \sum_{q} \left(1-e^{-i q} \right) \left(2 - \lambda_{qr} \right) \int_{0}^{t} \mathcal{\tilde{C}}^{\eta Q}_{q}(t,t) dt + \Gamma_{r}t. \nonumber \\
\end{eqnarray}
In the rhs of the above equation, $\mathcal{\tilde{C}}^{\eta Q}_q(t,t)$ in the first term is readily obtained from Eq.~\eqref{density_current_equal_soln}, while, in the second term, $\Gamma_r$ can be written in terms of gap distribition as given below,
\begin{eqnarray}
\label{gamma_r}
\hspace{-0.5 cm}
~~~\Gamma_{r} \hspace{0 cm}= \hspace{-0.05 cm} \rho \hspace{-0.1 cm} \sum_{l=\mid r \mid +1}^{\infty} \hspace{-0.25 cm} \phi(l) \hspace{-0.0 cm} \left[(l-\mid r \mid) \sum_{g=l}^{\infty}P(g)  \hspace{-0 cm}+ \sum_{g=\mid r \mid}^{l-1} (g-\mid r \mid) P(g) \right];~~~~
\end{eqnarray}
see Appendix E for calculation details. The quantity $\Gamma_{r}$ physically corresponds to the strength of the ``noise current" and is shown to be equal to the  two-point space-time correlation of the fluctuating current, i.e., $\langle J^{(fl)}_r(t) J^{(fl)}_0(0)\rangle = \Gamma_{r} \delta(t)$ [see Eq. \eqref{fluc_current_correlation}].
Later we also show that the steady-state fluctuation of the space-time integrated current $Q_{tot}(L,T) = \int_{0}^{T} dt\sum_{i=0}^{L-1}J^{(fl)}_i(t)$ [see Eq.~\eqref{I_fluctuation1}] satisfies, in the thermodynamic limit, a fluctuation relation,
\begin{eqnarray}
  \label{I_fluctuation}
2 \chi(\rho,\gamma) \equiv \lim_{L\rightarrow \infty} \frac{1}{LT} \langle Q_{tot}^{2}(L,T) \rangle = \sum_{r} \Gamma_{r} ,
\end{eqnarray}
where the particle mobility $\chi(\rho,\gamma)$ has the following expression,
\begin{eqnarray}\label{chi}
\chi(\rho,\gamma) = \frac{\rho}{2} \sum_{l=1}^{\infty}\phi(l) \left[\sum_{g=1}^{l-1}  g^{2}P(g) +  l^{2}\sum_{g=l}^{\infty}P(g)\right].
\end{eqnarray}
The mobility can be written explicitly as a function of density and tumbling rate, provided that the gap distribution $P(g)$ is known (see the $\rho, \gamma \rightarrow 0$ limit and the corresponding scaling regime, discussed later in Sec.~\ref{Sec:scaling_regime}).
Note that the sum rule, as in Eq.~\eqref{I_fluctuation}, states  that the scaled space-time integrated current fluctuation is equal to the spatially integrated correlation function for fluctuating current  and can be directly tested in simulations [see Fig.~(\ref{fig:space_time_cf})].

We now perform inverse Fourier transform of Eq.~\eqref{Q-Q_solution1} and finally obtain the desired solution for the steady-state unequal-time two-point current-current correlation function $\mathcal{C}^{QQ}_{r}(t',t)$ in real space,
\begin{widetext}
\begin{eqnarray}\label{Q1_Q2_solution_most_general}
\mathcal{C}^{QQ}_{r}(t',t) = -\frac{1}{2LD} \sum_{q}\frac{f_{q}}{\lambda_q^{2}}\left(e^{-\lambda_{q}Dt}- e^{-\lambda_{q}Dt'}\right)\left(e^{-\lambda_{q}Dt}- 1\right) e^{-iqr} - \frac{1}{2L}\sum_{q}\frac{f_{q}}{\lambda_q}\Bigg\{t - \frac{\left(1 - e^{-\lambda_{q}Dt} \right)}{\lambda_{q}D}  \Bigg\}(2-\lambda_{qr})  +\Gamma_{r} t. \nonumber \\
\end{eqnarray}
\end{widetext}
Now onwards, to keep the notations simple, we drop the argument of $D(\rho, \gamma)$ in Eq.~\eqref{Q1_Q2_solution_most_general} and elsewhere.

\subsection{Spatio-temporal correlation of the instantaneous current}
\label{Sec:space-time_inst}

In this section, we calculate the two-point unequal-time correlation function of the instantaneous current, i.e., $\mathcal{C}^{JJ}_{r}(t',t)$ in the steady state. We do this by differentiating the steady-state integrated current correlation function $\mathcal{C}^{QQ}_{r}(t',t)$ with respect to times $t'$ and $t$. However, the expression for $\mathcal{C}^{QQ}_{r}(t',t)$ provided in Eq.~\eqref{Q1_Q2_solution_most_general} is valid only for $t' \geq t$. Therefore, in order to obtain $\mathcal{C}^{JJ}_{r}(t',t)$ for arbitrary values of $t'$ and $t$, the appropriate formula can be written by using the Heaviside-Theta function $\Theta (t)$,
\begin{eqnarray}\label{inst_current_correlation_formula}
\mathcal{C}^{JJ}_{r}\hspace{-0.075 cm}(t,t')\hspace{-0.075 cm}=\hspace{-0.075 cm}\frac{d}{dt}\hspace{-0.075 cm}\frac{d}{dt'}\hspace{-0.075 cm}\left[\mathcal{C}^{QQ}_{r}(t',t) \Theta(t'-t) \hspace{-0.075 cm} +  \hspace{-0.075 cm}\mathcal{C}^{QQ}_{r}(t,t') \Theta(t-t') \right], \nonumber \\
\end{eqnarray}
where $\mathcal{C}^{QQ}_{r}(t,t')$ is obtained directly from Eq.~\eqref{Q1_Q2_solution_most_general} by interchanging $t'$ and $t$. Using Eq.~\eqref{inst_current_correlation_formula}, we straightforwardly compute $\mathcal{C}^{JJ}_{r}(t',t)$. After doing some algebraic manipulations and setting $t'=0$, we eventually arrive at the following expression,
\begin{eqnarray}
  \label{inst_current_correlation_most_general}
\mathcal{C}^{JJ}_{r}(t,t') &=& \Gamma_{r} \delta(t-t') - \frac{D}{4L}\sum_{q}\left(2 - \lambda_{qr} \right) f_{q} e^{-\lambda_{q}D  \vert t - t' \vert} \nonumber \\ && \hspace{2.5 cm} \Big\{\Theta(t-t') +\Theta(t'-t)\Big\}. 
\end{eqnarray}
where $f_q$ is obtained by substituting the steady-state gap-distribution $P(g)$ in  Eq.~\eqref{f_{q}}. Clearly, $\mathcal{C}^{JJ}_{r}(t,t')$ can be divided into two distinct parts. The space-dependent prefactor in the first component, which is delta correlated in time, pertains to the equal-time two-point correlation, which is determined by the fluctuating current correlations $\Gamma_r$. The second part comprises of the correlations at different space and time points. In the subsequent analysis, we examine the contribution of each of these terms.
\begin{figure*}
           \centering
         \includegraphics[width=0.33\linewidth]{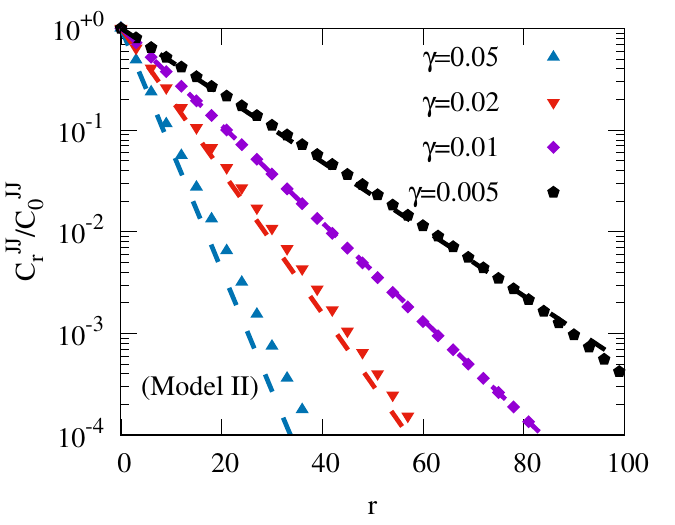}\hfill
         \includegraphics[width=0.33\linewidth]{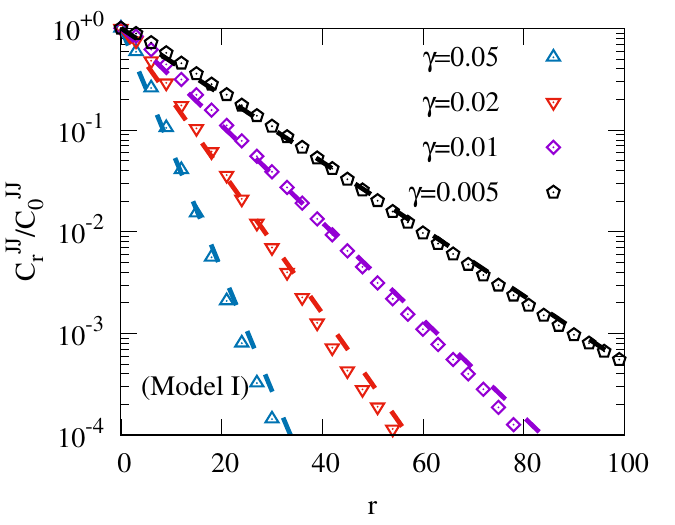}\hfill
         \includegraphics[width=0.33\linewidth]{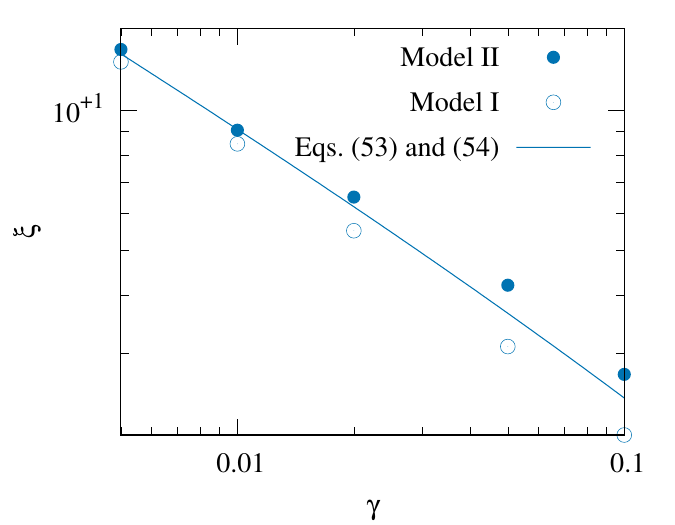}
         \caption{\textit{Verification of Eqs.~\eqref{inst_current_spatial_correlation_1}, \eqref{correlation_length} and \eqref{gap_size_large_persistence}-} We plot the scaled two point spatial correlation of the instantaneous current $\mathcal{C}^{JJ}_{r}/\mathcal{C}^{JJ}_{0}$ for model II (LLG, left-panel) and model I (standard RTPs, middle-panel), obtained from simulations (points), at a fixed $\rho=0.5$ and various $\gamma = 0.05$ (upper triangle), $0.02$ (lower triangle) and $0.01$ (diamond). We also compare the simulation data in both the models with the strong persistence analytical solution (dotted line) given by Eqs.~\eqref{inst_current_spatial_correlation_1}, \eqref{correlation_length} and \eqref{gap_size_large_persistence}. At the right panel, we plot the correlation length $\xi$, as a function of $\gamma$, at $\rho=0.5$ for model II (LLG, closed points) and model I (RTPs, open points) and compare them with the strong persistence analytical solution (line) provided by Eqs.~\eqref{correlation_length} and \eqref{gap_size_large_persistence}.}
          \label{fig:sp_correlation_inst_current}
\end{figure*}

\subsubsection{Equal-time unequal-space correlation}

To obtain the equal-time spatial correlations of the instantaneous current, we consider the case in Eq. (\ref{inst_current_correlation_most_general}) with $t=t'=0$, yielding the leading order contribution,
\begin{eqnarray}
  \label{inst_current_spatial_correlation}
\mathcal{C}^{JJ}_{r} \simeq  \left( \frac{\mathcal{C}^{JJ}_{0}}{\Gamma_0} \right) \Gamma_{r},
\end{eqnarray}
Note that the spatial dependence of the correlation function $\mathcal{C}^{JJ}_{r}$ is solely governed by $\Gamma_r$, which can be written in terms of the steady-state gap distribution function $P(g)$ only [see Eq.~\eqref{gamma_r}]. So, perhaps not surprisingly, the spatial correlations of current is in fact governed by the gap statistics and one would expect the correlation length to be determined by the typical gap size in the system. However, obtaining $P(g)$ explicitly as a function of $\rho$ and $\gamma$ is not an easy task in general. We can still do the following  asymptotic analysis, which is quite general. For larger gap size, we  expect $P(g)$ to be an exponential function (which can be shown to be indeed the case for $\gamma \ll 1$ \cite{subhadip_PRE_2021}),
\begin{eqnarray}\label{gap_dist_fn}
P(g) \simeq N_{*} \exp(-g/g_*),
\end{eqnarray}
where $g_*$ is the typcal gap size. Now, using the conservation relation $\langle g \rangle = \sum_{g=1}^{\infty}g P(g) =1/\rho -1$, one can immediately obtain the prefactor 
\begin{eqnarray}\label{prefactor}
N_{*} \simeq \left(\frac{1}{\rho}-1\right) \frac{(e^{1/g_*}-1)^{2}}{e^{1/g_*}}.
\end{eqnarray}
Upon substituting the above expression of $P(g)$ into Eq.~\eqref{gamma_r}, we  obtain the following simplified expression
\begin{eqnarray}\label{gamma_r_1}
\Gamma_r \simeq (1-\rho) \frac{(e^{1/g_*}-1)}{(e^{(\gamma + 1/g_*)}-1)}e^{-r/\xi} = \Gamma_{0}e^{-r/\xi},
\end{eqnarray}
which immediately leads to the spation correlation function of current,
\begin{eqnarray}\label{inst_current_spatial_correlation_1}
\mathcal{C}^{JJ}_{r} = \mathcal{C}^{JJ}_{0}e^{-r/\xi},
\end{eqnarray}
where the correlation length $\xi$ is given by
\begin{eqnarray}\label{correlation_length}
\xi=\frac{1}{\gamma + g_*^{-1}}.
\end{eqnarray}
Eqs.~\eqref{gamma_r_1} and \eqref{correlation_length} implies that the typical gap size $g_*$ plays a crucial role in determining $\Gamma_r$ and $\xi$. Although one can calculate $g_*$ numerically without much difficulty, determination of its exact analytical form for any arbitrary parameter regime is a nontrivial task. However, in the limit of strong persistence where $l_p = \gamma^{-1} \rightarrow \infty$, there is an analytic expression for typical gap size  \cite{subhadip_PRE_2021},
\begin{eqnarray}\label{gap_size_large_persistence}
 g_* \simeq \sqrt{\frac{1-\rho}{\gamma \rho }},
\end{eqnarray} 
which leads to the explicit solution of $\Gamma_r$ and hence the correlation function $C_r^{JJ}$. It is worth noting that, in this specific regime of strong persistence, the correlation length $\xi$ is primarily dominated by $g_*$ alone, which immediately implies $\xi \sim 1/\sqrt{\gamma}=\sqrt{\tau_p}$. That is, correlation length $\xi \sim \sqrt{\tau_p}$ diverges with persistence time $\tau_p$, thus providing a straightforward theoretical explanation of the findings in recent simulations and experiments \cite{Caprini_PRL_2020, Bertin_NatComm2020}.

In order to verify the theoretical results Eqs.~\eqref{inst_current_spatial_correlation_1}, \eqref{correlation_length} and \eqref{gap_size_large_persistence},
 in simulations we actually calculate, for better statistics, correlation function $C^{\bar{J} \bar{J}}(r) = \lim_{t \rightarrow \infty} \langle \bar{J}_0(t) \bar{J}_r(t) \rangle$ for a coarse-grained current $\bar{J}_i(t) = (1/\Delta t) \int_t^{t+\Delta t} dt J_i(t)$, averaged over a reasonably small time window $(t,t+\Delta t)$ and evaluated at two spatial points separated by distance $r$. 
In Fig.~(\ref{fig:sp_correlation_inst_current}), we plot the scaled correlation function $\mathcal{C}^{JJ}_{r}/\mathcal{C}^{JJ}_{0}$ for models II (LLG, closed points; left-panel) and I (standard RTPs, open points; middle-panel), obtained from Monte Carlo simulation, at different tumbling rates $\gamma=0.05$ (upper triangle), $0.02$ (lower triangle), $0.01$ (diamond) and $0.005$ (pentagon) while keeping the density fixed at $\rho=0.5$.
 We also compare the simulation data with the strong-persistence analytical solution (dotted lines), obtained using Eqs.~\eqref{inst_current_spatial_correlation_1}, \eqref{correlation_length} and \eqref{gap_size_large_persistence}. We indeed find a quite good agreement between simulations and analytic results in the limit of small $\gamma$. Finally, in the right panel of Fig.~(\ref{fig:sp_correlation_inst_current}), we plot the numerically obtained correlation length $\xi$ as a function of $\gamma$ for models II (LLG, closed points) and I (standard RTPs, open points) at a fixed density $\rho=0.5$ and also compare the results with the strong-persistence analytical solution, obtained using Eqs.~\eqref{correlation_length} and \eqref{gap_size_large_persistence}. In both cases - models I and II, we find in simulations that the correlation functions decay exponentially, $C^{\bar{J} \bar{J}}(r) \sim \exp(-r/\xi)$ and agree reasonably well with the analytical results.  
Notably, at small $\gamma$, the correlation lengths for models I and II asymptotically converge to each other as implied by our theory Eqs.~\eqref{inst_current_spatial_correlation_1}, \eqref{correlation_length} and \eqref{gap_size_large_persistence}.

\subsubsection{Equal-space unequal-time correlation function}

To evaluate the dynamic two-point correlation function for the instantaneous current in the steady state, we set $r=0$ and consider the case $t'=0$ and $t > 0$ in Eq.~\eqref{inst_current_correlation_most_general}, leading to the following expression,
\begin{eqnarray}\label{inst_current_temp_correlation_most_general}
\mathcal{C}^{JJ}_{0}(t, 0)=-\frac{D}{2L}\sum_{q}f_{q}(t) e^{-\lambda_{q}D t}.
\end{eqnarray}
\begin{figure}[tpb]
           \centering
         \includegraphics[width=0.9\linewidth]{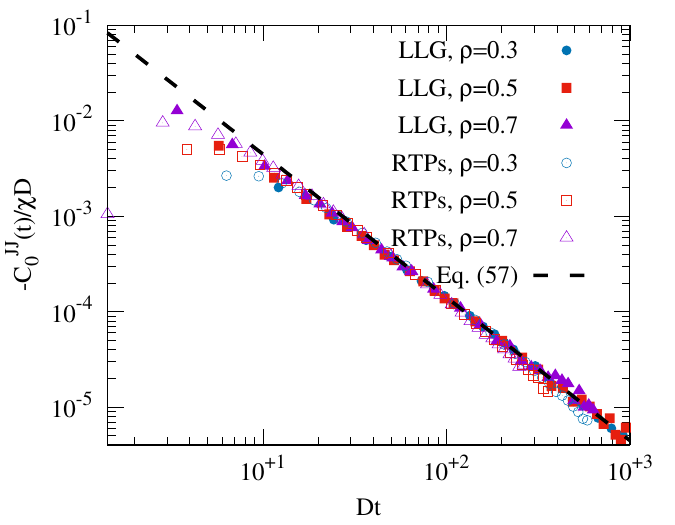}
         \caption{\textit{Verification of Eq.~\eqref{inst_current_temp_correlation_scaling}-} We plot the negative scaled equal-space unequal-time instantaneous current correlation function i.e. $-\mathcal{C}^{JJ}_{0}(t)/\chi D$, as a function of $Dt$, for model II (LLG, closed symbols) and model I (standard RTPs, open symbols), at a fixed $\gamma=0.1$ and different $\rho=0.3$ (blue circle), $0.5$ (red square) and $0.7$ (magenta triangle). We also compare the scaled simulation data points with the theoretical prediction (black dotted line) as shown in Eq.~\eqref{inst_current_temp_correlation_scaling}.}
          \label{fig:current_temp_correlation_scaled}
 \end{figure}
It is important to distinguish between the order of space and time limits. In the case, where the long-time limit $t \rightarrow \infty$ is taken  first and then the large system size limit $L \rightarrow \infty$ (i.e., $t \gg L^{2}/D$), it can be shown from Eq.~\eqref{inst_current_temp_correlation_most_general} that $\mathcal{C}^{JJ}_{0}(t, 0)= 0$. In the other case, where we first take the limit $L \rightarrow \infty$ and then the $t \rightarrow \infty$ limit (i.e., $L^{2}/D \gg t \gg 1/D$), we observe that the dynamic correlation$\mathcal{C}^{JJ}_{0}(t, 0)$ exbecomes a long-ranged one, which we characterize next. 
In this time regime, the behavior of $\mathcal{C}^{JJ}_{0}(t, 0)$ in Eq.~\eqref{inst_current_temp_correlation_most_general} is primarily dominated by the relaxations of the small-$q$ Fourier modes. Therefore, in order to obtain the large $t$ behavior, we can employ a small $q$ analysis by performing the transformations $\lambda_{q} \rightarrow q^{2}$ and $f_{q} \rightarrow \chi(\rho,\gamma) q^{2}$. Moreover, for large $L \gg 1$, by converting the summation into an integral, we obtain 
\begin{eqnarray}\label{inst_current_temp_correlation_asymptotic}
\mathcal{C}^{JJ}_{0}(t) \simeq - \frac{\chi(\rho,\gamma)}{4\sqrt{\pi D(\rho,\gamma)}} t^{-3/2};
\end{eqnarray}
see Appendix F for calculation details. Notably the correlation function $\mathcal{C}^{JJ}_{0}(t, 0)$ is {\it negatively correlated} for $t>0$, with a delta-functio at $t=0$, and exhibits a  power-law decay. These two important characteristics are a direct consequence of the observed subdiffusive growth in $\mathcal{C}^{QQ}_{0}(t, t)$ in the time regime $1 \ll Dt \ll L^{2}$. In fact, in this regime, upon suitable rearrangements of Eq.~\eqref{inst_current_temp_correlation_asymptotic}, we immediately obtain a scaling relation,
\begin{eqnarray}
  \label{inst_current_temp_correlation_scaling}
\frac{1}{\chi D}\mathcal{C}^{JJ}_{0}(t) = - \frac{1}{4\sqrt{\pi}}(D t)^{-3/2}.
\end{eqnarray} 
Here $\chi \equiv \chi(\rho,\gamma)$ and $D \equiv D(\rho, \gamma)$ are the density- and tumbling-rate-dependent collective mobility, as in Eq. \eqref{chi}, and bulk-diffusion coefficient, respectively. Interestingly, in a slightly different setting, dynamic fluctuations of ``force'' on a ``passive'' tracer immersed in a bath of hardcore RTPs has been studied in Ref. \cite{Kafri_2022_PRL}, where the associated correlation function was shown to have a similar power-law tail.

In order to verify the above scaling relationship in Eq. \eqref{inst_current_temp_correlation_scaling}, we first require to calculate the parameter dependent transport coefficients $D(\rho,\gamma)$ and $\chi(\rho,\gamma)$ for models I and II. We calculate $D(\rho,\gamma)$ and $\chi(\rho,\gamma)$ for model II (LLG) by numerically computing $P(g)$ in the steady-state and using it in Eqs.~\eqref{bulk-diffusivity-LH_sm} and \eqref{chi}. However, unlike model II, we do not have analytical expressions for the transport coefficients in model I (standard RTPs), forcing us to rely on direct simulations. To calculate $D(\rho,\gamma)$ for model I, we study the relaxation of long-wavelength perturbation and follow the method provided in our previous work \cite{Tanmoy_condmat2022}. For the numerical determination of the mobility $\chi(\rho,\gamma)$, we compute the scaled space-time integrated current fluctuation as given in Eq.~\eqref{I_fluctuation1}. We now verify the scaling relation in Eq.~\eqref{inst_current_temp_correlation_scaling} by plotting the ratio $-\mathcal{C}^{JJ}_{0}(t)/\chi D$ (obtained from simulations) in Fig.~(\ref{fig:current_temp_correlation_scaled}) as a function of  scaling variable $Dt$, for model II (closed points) and model I (open points) for various densities $\rho= 0.3$ (circle), $0.5$ (square) and $0.7$ (triangle) at a fixed tumbling rate $\gamma=0.1$. We also compare  simulation results (points) with the analytic solution as in Eq.~\eqref{inst_current_temp_correlation_scaling}. We found our theory to agree quite well with simulations at large times.

\subsection{Space-time correlations of fluctuating (``noise'') current}
\label{Sec:space-time_fluc}

In this section, we focus on determining the two-point spatio-temporal correlation function for fluctuating current $J^{(fl)}(t)$ in the steady state. In other words, our aim is to derive the expression for $\mathcal{C}^{J^{(fl)}J^{(fl)}}_{r}(t, 0)$ where $t \geq 0$. Using the current decomposition as in Eq. \eqref{current_decompose}, we can write the following relation,
\begin{eqnarray}\label{inst_current_correlation_decomposition}
\mathcal{C}^{J^{fl}J^{fl}}_{r}(t, 0)=\mathcal{C}^{JJ}_{r}(t, 0) - \mathcal{C}^{J^{D}J}_{r}(t, 0) - \mathcal{C}^{J^{fl}J^{D}}_{r}(t, 0). \nonumber \\
\end{eqnarray}
Notably, the fluctuation current $J^{(fl)}(t)$ at time $t$ is not correlated with the diffusive current $J^{D}_{0}(0)$ at an earlier time $t=0$, i.e.,
\begin{eqnarray}\label{fluc_diffusive_curr_correlation}
\mathcal{C}^{J^{fl}J^{D}}_{r}(t, 0) = \langle J^{fl}_{r}(t) J^{D}_{0}(0) \rangle =0.
\end{eqnarray}
Then the third term in Eq.~\eqref{inst_current_correlation_decomposition} immediately drops out. Moreover, in order to determine the second term $\mathcal{C}^{J^{D}J}_{r}(t, 0)$, we use the following relation
\begin{eqnarray}\label{diff_inst_correlation_1}
\mathcal{C}^{J^{D}J}_{r}(t, 0) &=&\left[\frac{d}{dt'}\mathcal{C}^{J^{D}Q}_{r}(t, t')\right]_{t'=0},  \\&\simeq& D\frac{d}{dt'} \left[\mathcal{C}^{\eta Q}_{r}(t, t')-\mathcal{C}^{\eta Q}_{r+1}(t, t') \right]_{t'=0},\label{diff_inst_correlation_2}
\end{eqnarray}
where we have used the truncation approximation as in Eq.~\eqref{closure_approximation}, to arrive  at Eq.~\eqref{diff_inst_correlation_2} by using  Eq.~\eqref{diff_inst_correlation_1}. Following Eq.~\eqref{Fourier_transform}, we now expand the correlators  $\mathcal{C}^{\eta Q}_{r}(t, t')$ in the Fourier basis, and then using Eq.~\eqref{density_current_unequal_soln}, we obtain the desired solution,
\begin{eqnarray}\label{diff_inst_correlation_solution}
\mathcal{C}^{J^{D}J}_{r}(t, 0)=-\frac{D}{4L}\sum_{q}(2-\lambda_{qr})f_{q}(t)e^{-\lambda_q Dt}.
\end{eqnarray}
Importantly the above solution coincides with the two point unequal-time correlation of $\mathcal{C}^{JJ}_{r}(t, 0)$ which is displayed in the second term of Eq.~\eqref{inst_current_correlation_most_general} with $t \geq t'=0$. Finally using Eqs.~\eqref{inst_current_correlation_most_general}, \eqref{fluc_diffusive_curr_correlation} and \eqref{diff_inst_correlation_solution} in Eq.~\eqref{inst_current_correlation_decomposition}, we readily obtain
\begin{eqnarray}\label{fluc_current_correlation}
\mathcal{C}^{J^{fl}J^{fl}}_{r}(t, 0)= \langle J^{fl}_{r}(t) J^{fl}_{0}(0) \rangle= \delta(t) \Gamma_{r}.
\end{eqnarray}
\subsection{Fluctuation of the space-time integrated current}\label{Sec:fluc_total}
The space-time integrated current $Q_{tot}(L,T)$ of the system is defined as
\begin{eqnarray}\label{space-time-integrated-current}
Q_{tot}(L,T)= \sum_{i=0}^{L-1}Q_{i}(T) = \int_{0}^{T}dt\sum_{i=0}^{L-1}J_{i}(t).
\end{eqnarray}
Note that, $Q_{tot}(L,T)$ measures the total current in the system upto the observtion time $T$. Alternatively, $Q_{tot}(L,T)$ can be linked to the cumulative tagged particle displacements in the following way:
\begin{eqnarray}\label{space-time-integrated-current_1}
Q_{tot}(L,T)= \sum_{i=1}^{N}X_{i}(T),
\end{eqnarray}
where $X_{i}(T)$ is the displacement of the $i$th particle in time $T$.  In this section, we will characterize the fluctuation properties of $Q_{tot}(L, T)$, which essentially calculates the fluctuation of cumulative displacements of active particles in the system. 

\begin{figure*}
           \centering
         \includegraphics[width=0.45\linewidth]{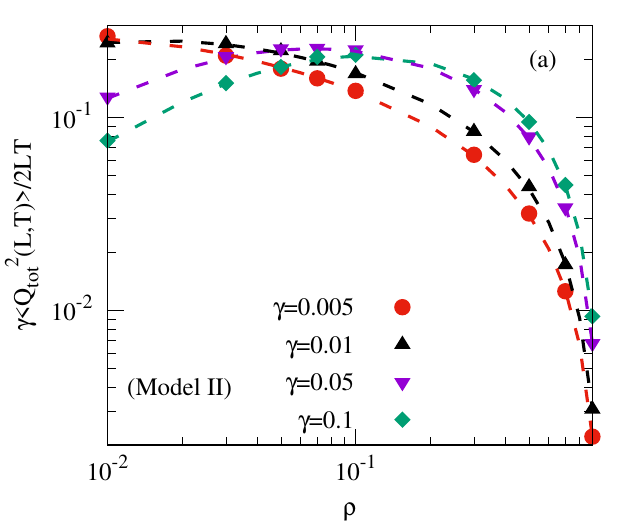}\hfill
         \includegraphics[width=0.45\linewidth]{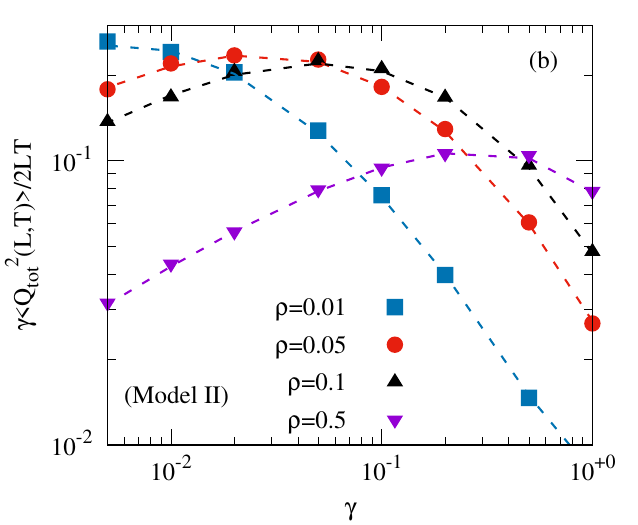}\vfill
         \includegraphics[width=0.45\linewidth]{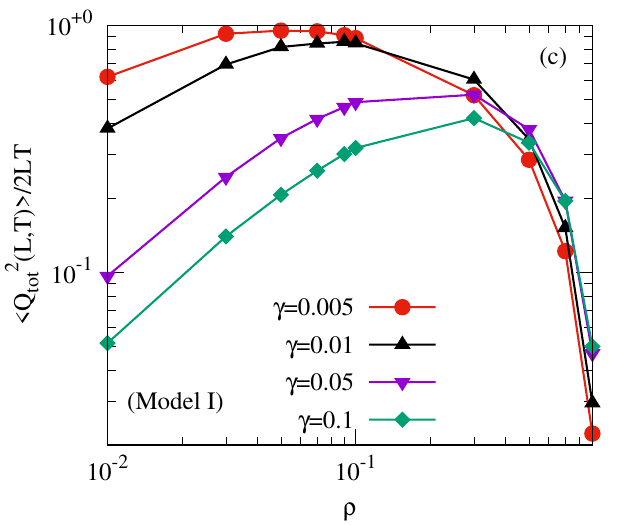}\hfill
         \includegraphics[width=0.45\linewidth]{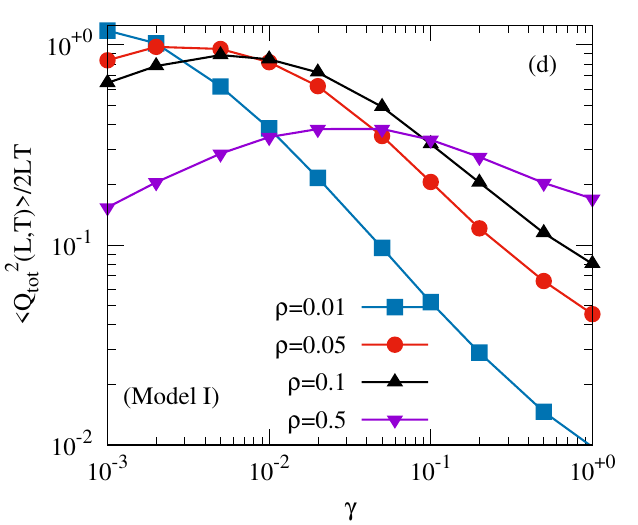}
         \caption{In panels (a) and (b), we plot the scaled space-time integrated current fluctuation $\gamma \langle Q_{tot}^{2}(L,T) \rangle/2LT$ for the LLG, obtained from simulation (points), as a function of $\rho$ [at different $\gamma=0.001$ (blue square), $0.005$ (red circle), $0.01$ (black upper-triangle), $0.05$ (magenta down-triangle), and $0.1$ (green diamond)] and $\gamma$ [at various $\rho=0.01$ (blue square), $0.05$ (red circle), $0.1$ (black upper-triangle) and $0.5$ (magenta down-triangle)], respectively. Corresponding dotted lines are $\gamma \chi(\rho, \gamma)$ calculated by using the numerically obtained $P(g)$ in Eq.~\eqref{chi}. The excellent match between these two quantities verifies Eq.~\eqref{I_fluctuation1}. In panels (c) and (d), we plot $\langle Q_{tot}^{2}(L, T) \rangle/2LT$ for model I (standard RTPs), obtained from numerical simulation (line-point), as a function of $\rho$ and $\gamma$, respectively for the aforementioned parameters.}
          \label{fig:space_time_cf}
 \end{figure*}

It is worth noting that, according to the decomposition shown in Eq.~\eqref{current_decompose}, we can decompose $J_{i}(t)$ in Eq.~\eqref{space-time-integrated-current} into diffusive $J^{(D)}_i(t)$ and fluctuating $J^{(fl)}_{i}(t)$ components. However, for the periodic system under consideration here, we use the identity
\begin{eqnarray}
\sum_{i=0}^{L-1}J^{(D)}_{i}(t)=0.
\end{eqnarray}
As a result, the diffusive component does not contribute to $Q_{tot}(L,T)$, leaving us with,
\begin{eqnarray}\label{space-time-integrated-current_2}
Q_{tot}(L,T)= \int_{0}^{T}dt\sum_{i=0}^{L-1}J^{fl}_{i}(t).
\end{eqnarray}
The above equation clearly reflects the fact that the total current of the system $Q_{tot}(L,T)$ is solely governed by the fluctuating component $J^{(fl)}_{i}(t)$, which immediately implies the average current
\begin{eqnarray}\label{total_current_avrg}
\langle Q_{tot}(L,T) \rangle = \int_{0}^{T}dt\sum_{i=0}^{L-1}\langle J^{fl}_{i}(t) \rangle = 0,
\end{eqnarray}
since $\langle J^{fl}_{i}(t) \rangle = 0$. In a similar way, we write the expression for the fluctuation
\begin{eqnarray}\label{total_current_fluc}
\langle Q^{2}_{tot}(L,T) \rangle &=& \int_{0}^{T}dt' \int_{0}^{T}dt\sum_{i=0}^{L-1} \sum_{r}\langle J^{fl}_{i}(t) J^{fl}_{i+r}(t') \rangle.\nonumber \\
\end{eqnarray}
Using Eq.~\eqref{fluc_current_correlation} in the above equation, it is now straightforward to find that the total current fluctuation satisfies the following relation
\begin{eqnarray}\label{I_fluctuation1}
\frac{1}{LT}\langle Q^{2}_{tot}(L,T)\rangle = \sum_{r} \Gamma_{r} = 2\chi(\rho,\gamma),
\end{eqnarray}
where we have already expressed $\chi(\rho,\gamma)$ in terms of $P(g)$ in Eq.~\eqref{chi}. To verify the above equation for the model II (LLG), we first compute $\langle Q^{2}_{tot}(L,T)\rangle$ from numerical simulation with $L=1000$ and $T=50$ in the parameter range $0.01 \leq \rho \leq 0.9$ and $0.005 \leq \gamma \leq 1$. Moreover, we also numerically compute $P(g)$ and use them in Eq.~\eqref{chi} to obtain $\chi(\rho,\gamma)$ for the same parameter values. In panels (a) and (b) of Fig.~(\ref{fig:space_time_cf}), we plot the numerically obtained scaled fluctuation $\gamma\langle Q^{2}_{tot}(L,T)\rangle/2LT$ as a function of $\rho$ and $\gamma$, respectively. We also plot the already calculated $\gamma \chi(\rho,\gamma)$ and represent them with the dotted lines. The excellent agreement between $\gamma\langle Q^{2}_{tot}(L,T)\rangle/2LT$ and $\gamma \chi(\rho,\gamma)$ immediately verifies Eq.~\eqref{I_fluctuation1} for model II. In panels (c) and (d) of Fig.~(\ref{fig:space_time_cf}), we plot the functional variation of the numerically obtained $\langle Q^{2}_{tot}(L,T)\rangle/2LT$, which is also a measure of $\chi(\rho, \gamma)$ according to Eq.~\eqref{I_fluctuation1}, with respect to $\rho$ and $\gamma$, respectively for model I (standard RTPs). We notice that the scaled fluctuation is nonmonotonic in both $\rho$ and $\gamma$, and we find qualitative similarities between panels (a) and (c), as well as panels (b) and (d), thus establishing the same underlying mechanism of particle transport in models I (standard RTPs) and II (LLG).  

\subsubsection*{Scaling regime for the particle mobility $\chi(\rho, \gamma)$}\label{Sec:scaling_regime}

In this section, we study the particle mobility $\chi(\rho, \gamma)$ in the two limiting cases: \textit{Case I, $\rho \rightarrow 0$, $\gamma \rightarrow \infty$} and \textit{Case II, $\rho \rightarrow 0$, $\gamma \rightarrow 0$}.  
While Case I is qualitatively similar to the SSEP limit, Case II remarkably captures the nontrivial interplay of interaction and persistence in terms of the scaling variable $\psi=\rho/\gamma$. Previously, in Ref. \cite{Tanmoy_condmat2022}, we investigated the scaling regime for the bulk-diffusion coefficient $D(\rho,\gamma)$ for Case II; also, in this case, we calculated the associated scaling function analytically for model II. By using the truncation scheme as in Eq. \eqref{closure_approximation}, we have been able to calculate the same for the mobility $\chi(\rho,\gamma)$.

By using Eqs.~\eqref{gap_dist_fn} and \eqref{chi}, we substitute the steady-state gap distribution $P(g)$ in $\chi(\rho,\gamma)$ for model II and obtain
\begin{eqnarray}\label{chi-1}
\chi(\rho,\gamma) \simeq \frac{(1-\rho)}{2}\frac{(e^{1/g_*}-1)(e^{\gamma + 1/g_*}+1)}{(e^{\gamma + 1/g_*}-1)^{2}}.
\end{eqnarray}
Note that, the above expression of $\chi(\rho,\gamma)$ is valid for any arbitrary $\rho$ and $\gamma$. However, in the following discussion, we analyze the limiting cases mentioned in the beginning.
\\
\\
{\it Case I, $\rho \rightarrow 0$ and $\gamma \rightarrow \infty$.--} In this case of small persistence and low density limit, the steady-state distribution is a product measure: $P(g) \sim (1-\rho)^{g} \simeq e^{-\rho g}$, yielding $g_*=1/\rho$.  Finally, using this $g_*$ and setting the condition $\gamma \gg 1 \gg \rho$ in Eq.~\eqref{chi-1}, we obtain
\begin{eqnarray}\label{chi_ssep_limit}
\chi(\rho,\gamma) \simeq  \frac{e^{-\gamma}}{2} \rho(1-\rho) =\frac{e^{-\gamma}}{2} \chi^{(0)},
\end{eqnarray}
which, upto a scale factor $\exp(-\gamma)$ due to time scaling (explained below), is the particle mobility  $\chi^{(0)}=\rho(1-\rho)$ in SSEP.
The exponential prefactor $e^{-\gamma}$ in the above equation appears because of the following. In this case, $\phi(l)$ carries maximum weight at $l=0$, while all other hop-lengths, i.e., $l >0$,  occur with an exponentially smaller probability $1-\phi(0)=e^{-\gamma}$. Furthermore, among these nonzero hop-lengths, $l=1$ dominates, with contributions from the larger $l$ being vanishingly small. Therefore, in the limit $\gamma \rightarrow \infty$ or, equivalently, $l_p \rightarrow 0$ in model II (LLG), particles effectively perform SSEP-like hopping dynamics, albeit with a rate that is simply exponentially small, $e^{-\gamma}$, thus explaining the prefactor in Eq.~\eqref{chi_ssep_limit}.  
\\
\\
{\it Case II, $\rho \rightarrow 0$ and $\gamma \rightarrow 0:$} As we show below, similar to the bulk-diffusion coefficient $D(\rho,\gamma)$ in hardcore RTPs studied in Ref. \cite{Tanmoy_condmat2022}, the particle mobility too exhibit interesting scaling properties. Indeed, in the strong-persistence and low-density limit, there are  only two relevant length scales in the problem: The persistence length  $l_p=v/\gamma$ and the ``mean free path" or the average gap $\langle g \rangle \simeq 1/\rho$. Consequently their ratio  $\psi = l_p / \langle g \rangle$ is expected to provide a scaling variable that should quantify the interplay of persistence and interaction. In the regime of strong persistence, $\psi \rightarrow \infty$ denotes the strongly interacting limit, whereas $\psi \rightarrow 0$ corresponds to the noninteracting limit. Now as argued previously in Ref. \cite{Tanmoy_condmat2022}, we have the typical gap-length $g_*$ satisfing the following scaling law - $g_* \simeq \mathcal{G}(\psi)/\rho$. In the limit of $\psi\rightarrow 0$, the model reduces to the well known SSEP for which $\mathcal{G}(\psi)=1$, while, in the opposite limit of $\psi \rightarrow \infty$, we have the strongly interacting regime, for which previous calculations in Ref.~\cite{subhadip_PRE_2021} yield $\mathcal{G}(\psi)=\sqrt{\psi}$. Now combining these two limiting cases, we could simply write $\mathcal{G}(\psi) \simeq \sqrt{1+\psi}$. Finally, plugging the assumed form of $g_* ={\cal G}(\psi)/\rho$ in Eq.~\eqref{chi-1}, and after some algebraic manipulations, we obtain the following scaling law,
 \begin{eqnarray}
  \chi_{LLG}(\rho, \gamma) \equiv \frac{\chi^{(0)}}{ \gamma^{2}} \mathcal{H}_{LLG}\left(\psi=\frac{\rho}{\gamma}\right),
       \label{transport_scaling}
     \end{eqnarray}
where $ \chi^{(0)} = \rho(1-\rho)$ is the particle mobility in the SSEP and the expression for the scaling function can be explicitly written as
 \begin{eqnarray}\label{scaling_functions1}
  \mathcal{H}_{LLG}(\psi) &=& \frac{\mathcal{G}(\psi)}{(\psi + \mathcal{G}(\psi))^{2}}.
 \end{eqnarray}
 Finally for model II (LLG), by replacing the above form of $\mathcal{G}(\psi) \simeq \sqrt{1+\psi}$, we immediately obtain
 \begin{eqnarray}\label{scaling_functions}
  \mathcal{H}_{LLG}(\psi) &=& \frac{\sqrt{1+\psi}}{(\psi + \sqrt{1+\psi})^{2}};
 \end{eqnarray}
 for calculation details, see Appendix I.

\begin{figure}
           \centering
         \includegraphics[width=0.9\linewidth]{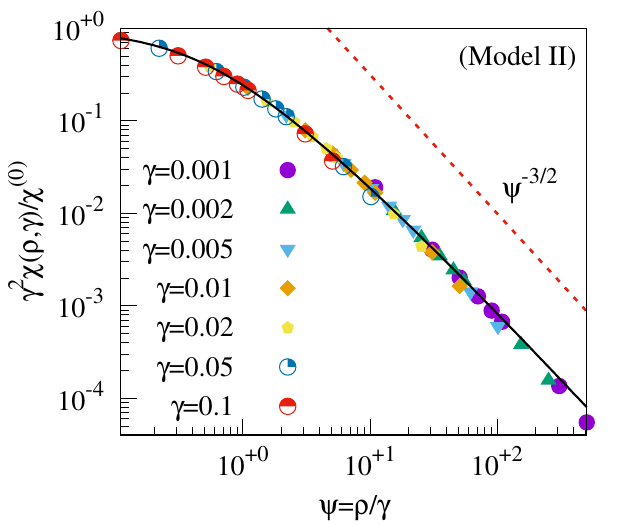}
         \includegraphics[width=0.9\linewidth]{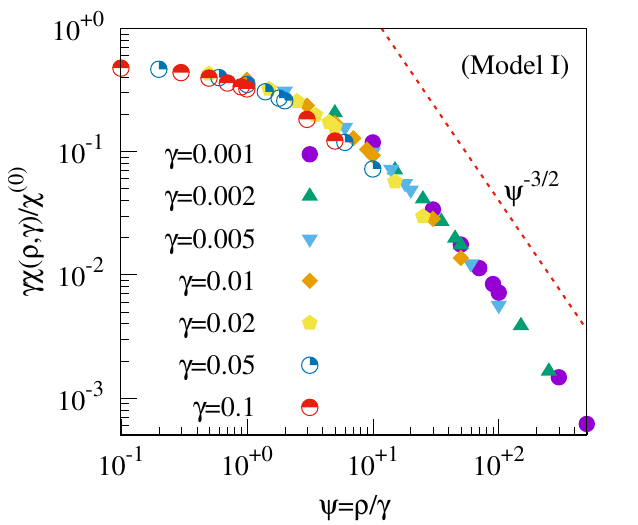}
         \caption{\textit{Verification of Eqs.~\eqref{transport_scaling} and \eqref{scaling_functions}-} We plot the ratio $\gamma^{a}\chi(\rho,\gamma)/\chi^{(0)}$ for model II (LLG, top-panel, $a=2$) and model I (RTPs, bottom-panel, $a=1$), as a function of scaling variable $\psi=\rho/\gamma$ in the parameter ranges $0.01 \leq \rho \leq 0.5$ and $0.001 \leq \gamma \leq 0.1$. For LLG, we compare the collapsed simulation data points with the analytic scaling function  $\mathcal{H}_{LLG}(\psi)$ (solid balck line) shown in Eq.~\eqref{scaling_functions}. For both the models, the collapsed data points exhibit $\psi^{-3/2}$ decay in the asymptotic limit, which is shown here by the red-dotted line.}
          \label{fig:transport_coefficients_scaled}
\end{figure}

In top-panel of Fig.~\ref{fig:transport_coefficients_scaled}, we plot the scaled mobility $\gamma^{2}\chi_{LLG}(\rho,\gamma)/\chi^{(0)}$, obtained from simulation (points), as a function of the scaling variable $\psi=\rho/\gamma$ in the parameter ranges $0.01 \leq \rho \leq 0.5$ and $0.001 \leq \gamma \leq 0.1$. We observe the data points collapse with each other, and that the collapsed data points follow the analytically obtained scaling function $\mathcal{H}_{LLG}(\psi)$ (solid line) very well. This observation indeed verifies the existence of the scaling law as in Eq. \eqref{transport_scaling} and  substantiate the scaling function in Eq.~\eqref{scaling_functions}. We find that the same scaling law holds also for model I (standard RTPs). To demonstrate this, we plot $\gamma\chi_{RTP}(\rho,\gamma)/\chi^{(0)}$ as a function of $\psi$ in the bottom-panel of Fig.~\ref{fig:transport_coefficients_scaled}, in the same parameter range as LLG.

\subsection{Time-Integrated Bond-Current fluctuation}\label{Sec:bond_current_fluc}

In order to calculate the steady-state time-integrated bond-current fluctuation, we simply put $r=0$ and $t'=t=T$ in Eq.~\eqref{Q1_Q2_solution_most_general} and obtain
\begin{eqnarray}\label{bond-current-fluc-0}
\hspace{-0.5 cm}\mathcal{C}^{QQ}_{0}(T,T) &=&\Gamma_{0} T - \frac{1}{L} \sum_{q}\frac{f_q}{\lambda_q} \Bigg\{t - \frac{\left(1 - e^{-\lambda_{q} DT} \right)}{\lambda_{q} D}  \Bigg\}, \\ \label{bond-current-fluc-1}
&=&\frac{2 \chi}{L} T + \frac{1}{DL}\sum_{q}\frac{f_{q}}{\lambda_q^{2}} \left(1 - e^{-\lambda_{q} DT} \right);
\end{eqnarray}
see Appendix G for the derivation of Eq.~\eqref{bond-current-fluc-1}. It is important to mention that $\mathcal{C}^{QQ}_{0}(T,T)$, which is expressed by Eqs.~\eqref{bond-current-fluc-0} and \eqref{bond-current-fluc-1} in its most comprehensive form, exhibits quite interesting characteristics at different time regimes. In the subsequent discussion, we investigate these properties by analyzing the limiting cases.

\subsubsection{Case I : Small-time regime $DT \ll 1$}

In this case, we perform a linear expansion of the exponential function in the second term of Eq.~\eqref{bond-current-fluc-0} with respect to $DT$, which yields $e^{-\lambda_{q} DT} \approx 1 - \lambda_{q} DT$. As a result, this term drops out and we are left with
\begin{eqnarray}\label{bond-current-fluc-smallest}
\mathcal{C}^{QQ}_0 (T,T) \simeq \Gamma_{0} T,
\end{eqnarray}
where $\Gamma_{0}$ is simply obtained by putting $r=0$ in Eq.~\eqref{gamma_r}.

\subsubsection{Case II : Intermediate- and long-time regime $DT \gg 1$}

\begin{figure}[tpb]
           \centering
         \includegraphics[width=0.9\linewidth]{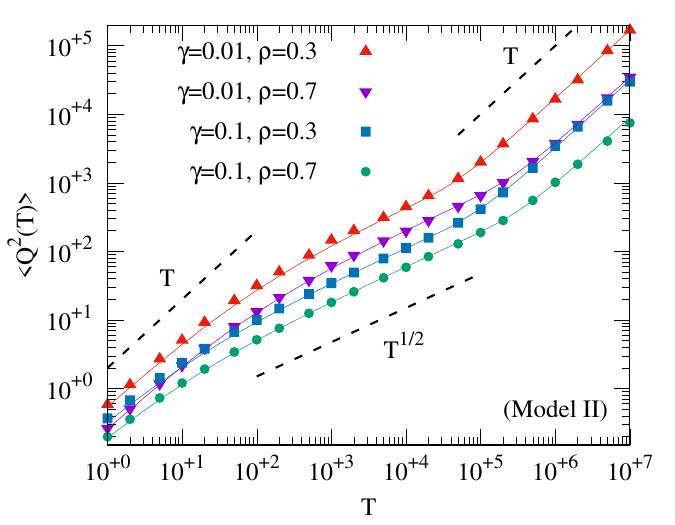}\vfill
         \includegraphics[width=0.9\linewidth]{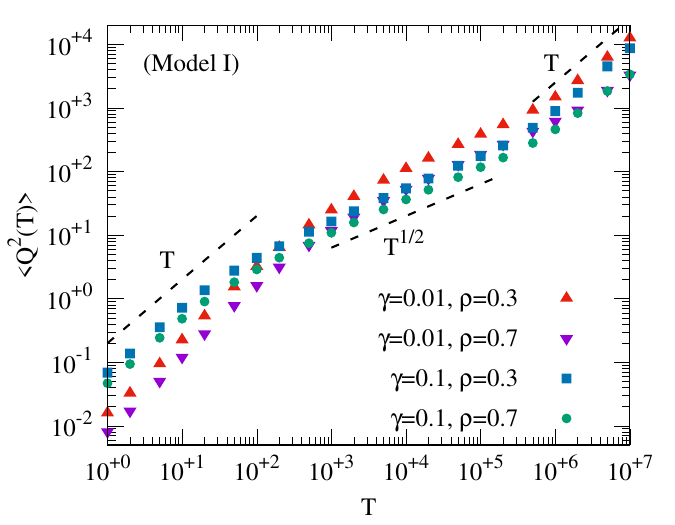}
         \caption{We plot the time-integrated bond-current fluctuation $\langle Q_{i}^{2}(T) \rangle $, as a function of time $T$, obtained from simulations (points) for model II (LLG, top-panel) and model I (standard RTPs, bottom-panel) at $\rho=0.3,$ $0.7$ and $\gamma=0.1$, $0.01$. In case of model I, we also compare the simulation data points with the analytical solution shown in Eq.~\eqref{bond-current-fluc-1} (line). For both these models, $\langle Q_{i}^{2}(T) \rangle $ exhibits subdiffusive growth at the intermediate time regime, followed by a diffusive (linear) growth, as shown by the dotted lines. However, the behavior at the very small time is model dependent; for model II (LLG), it is always diffusive (linear), but for model I (standard RTPs), it depends on the parameter values. It exhibits superdiffusive growth at smaller $\gamma=0.01$ and diffusive growth at larger $\gamma=0.1$.}
          \label{fig:current_fluctuation_unscaled}
 \end{figure}

In general, it is difficult to solve the summation in the second term of Eq.~\eqref{bond-current-fluc-1}. However, in this case, note that the summand only contributes when $q \rightarrow 0$ and vanishes otherwise. In this limit, the eigenvalues are quadratic, i.e. $\lambda_{q} \rightarrow q^{2}$, $\lambda_{gq} \rightarrow g^{2}q^{2}$ and $\lambda_{lq} \rightarrow l^{2}q^{2}$, resulting in a simplified version of Eq.~\eqref{bond-current-fluc-1},
\begin{eqnarray}
\label{bond-current-fluc-2}
\mathcal{C}^{QQ}_{0}(T,T)=\frac{2 \chi}{L} \left[T + \frac{1}{D} \sum_{q} \frac{1}{\lambda_q^{2}} \left(1 - e^{-\lambda_{q} D T}\right) \right].
\end{eqnarray}
Considering both of the preceeding cases, the limiting behavior of   $\mathcal{C}^{QQ}_{0}(T,T)=\langle Q^{2}(t)\rangle$ is obtained to be
\begin{eqnarray} 
 \langle Q^{2}(T) \rangle \simeq 
\left\{
\begin{array}{ll}
\vspace{0.15 cm}
  \Gamma_{0} T,            ~~~  & {\rm for}~   DT \ll 1, \\
\vspace{0.15 cm}
  \frac{2\chi(\rho,\gamma)}{\sqrt{\pi D(\rho, \gamma)}}\sqrt{T},            ~~~  & {\rm for}~  1 \ll DT \ll L^2, \\
\vspace{0.15 cm}
 \frac{2\chi(\rho,\gamma)}{L} T,            ~~~  & {\rm for}~ DT \gg L^{2}, \\
\end{array}
\right.
\label{cf_limit}
\end{eqnarray}
where the first term simply corresponds to Eq.~\eqref{bond-current-fluc-smallest}, while the other two are obtained using Eq.~\eqref{bond-current-fluc-2}; for details, see Appendix H. Therefore, the time-integrated bond-current fluctuation $\langle Q_{i}^{2}(T) \rangle$ exhibits an initial linear growth in time  before transitioning to a subdiffusive scaling at intermediate regime $ L^{2}/D \gg T \gg 1/D$, subsequently, at larger time scales $T \gg L^{2}/D$, it reverts to diffusive or linear scaling behavior where the strength of the fluctuation is governed by $\chi(\rho,\gamma)$. In order to verify these observations, we plot the numerically obtained bond-current fluctuation $\langle Q_{i}^{2}(T)\rangle$, as a function of the observation time $T$, at the top-panel of Fig.~(\ref{fig:current_fluctuation_unscaled}) for LLG at different parameter values and compare them with our analytical solution given in Eq.~\eqref{bond-current-fluc-1}. We find the numerical data points clearly exhibit previously mentioned three different regimes and it follows the analytical solution quite well. In case of model I (standard RTPs), we plot the similar quantity, obtained from simulation, at the bottom-panel of Fig.~(\ref{fig:current_fluctuation_unscaled}). We find the data points to follow similar characteristics with model II (LLG) in the time regime much larger than its microscopic time-scale i.e. when $DT \gg 1$. However, in the other limit of smaller time scale, RTPs exhibit superdiffusive behavior, i.e. $\langle Q_{i}^{2}(T)\rangle \sim T^{\alpha}$ where $\alpha > 1$ is parameter dependent.\\ 
\begin{figure*}
           \centering
         \includegraphics[width=0.33\linewidth]{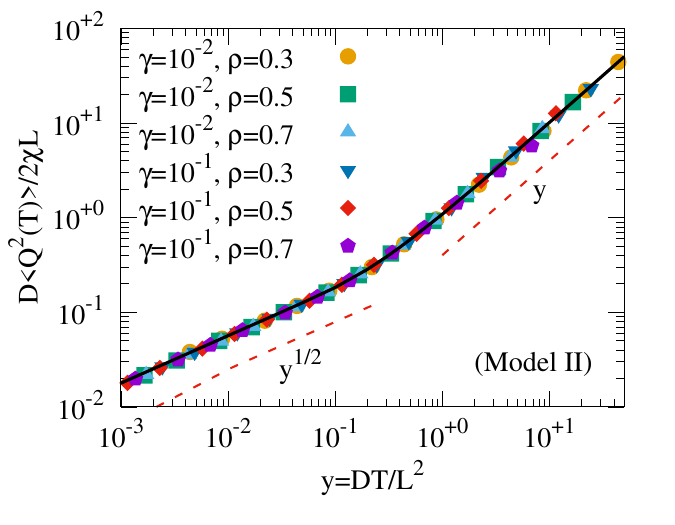}\hfill
         \includegraphics[width=0.33\linewidth]{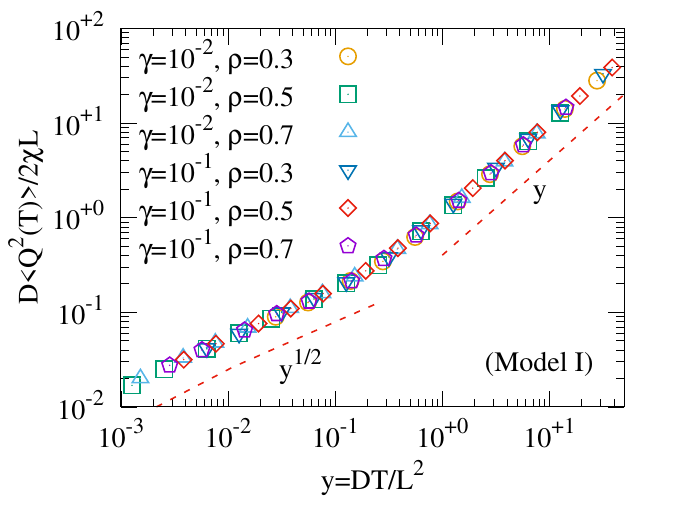}\hfill
         \includegraphics[width=0.33\linewidth]{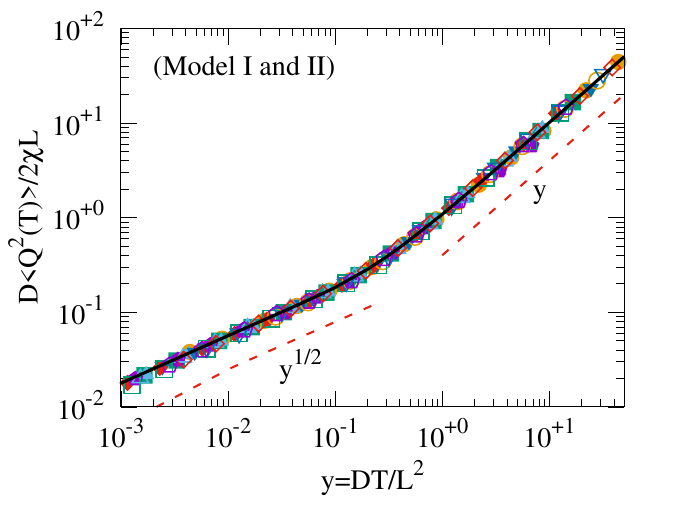}
         \caption{\textit{Verification of Eqs.~\eqref{bond-current-fluc-3} and \eqref{cf_scaling_function}-} We plot the scaled bond-current fluctuation $D\langle Q_{i}^{2}(T) \rangle/2\chi L$ for model II (LLG, left-panel) and model I (standard RTPs, middle-panel), obtained from simulations (points) at various $\rho$ and $\gamma$, as a function of the rescaled hydrodynamic time $y=D(\rho,\gamma)T/L^{2}$. For LLG, we compare the scaled data points with the analytic scaling function $\mathcal{W}\left(y\right)$ shown in Eq.~\eqref{cf_scaling_function} (black line). In the right-panel, we check the universality of $\mathcal{W}\left(y\right)$ by plotting these numerically obtained scaled current fluctuation $D\langle Q_{i}^{2}(T) \rangle/2\chi L$ for model II (LLG, closed points) and model I (standard RTPs, open points) together and compare them with the analytically obtained $\mathcal{W}\left(y\right)$. In all three panels, the red dotted guiding lines reflect the early time subdiffusive $(\sim \sqrt{y})$, followed by the diffusive growth of $\mathcal{W}\left(y\right) \sim y$ as derived in Eq.~\eqref{cf_scaling_function_limit}.}
          \label{fig:current_fluctuation_scaled}
 \end{figure*}

Interestingly, the above intermdetiate and long time regime for time-dependent bond-current fluctuations can actually be unified through a single scaling function as following. Moreover, quite remarkably, the scaling function seems to be universal as it does not dependend on the details of the dynamical rules of the models considered here. In the limit of $L \rightarrow \infty$ and $DT \rightarrow \infty$ such that $y=DT/L^{2}$ is finite, we find $\mathcal{C}^{QQ}_{0}(T,T) = \langle Q^{2}(T) \rangle$, as expressed in Eq.~\eqref{bond-current-fluc-2}, to satisfy the following scaling relation,
\begin{eqnarray}\label{bond-current-fluc-3}
 \frac{D}{2\chi L} \langle Q^{2}(T) \rangle = \mathcal{W}\left(\frac{DT}{L^{2}}\right).
\end{eqnarray} 
For model II (LLG), the scaling function is calculated exactly within the truncation scheme and is is given by the following series,
\begin{eqnarray}\label{cf_scaling_function}
\mathcal{W}\left(y\right) =y + \lim_{L \rightarrow \infty} \frac{1}{L^{2}}\sum_{q}\frac{1}{\lambda_q}\left(1 - e^{-\lambda_{q}yL^{2}} \right).
\end{eqnarray}
One can approximate the discrete sum in the rhs of the above equation as a quadrature, which can be written in terms of  known functions  \cite{Hazra_2023},
\begin{eqnarray}
\mathcal{W}\left(y\right) \simeq y + \left( \frac{y}{\pi} \right)^{1/2} {\rm erfc} \left(2 \pi \sqrt{y} \right) + \frac{1-\exp(-4 \pi^2 y)}{4 \pi^2}, ~~~
\end{eqnarray}
with ${\rm erfc}(y) = 1- {\rm erf}(y)$ and the error function ${\rm erf}(y) = (2/\sqrt{\pi}) \int_0^y \exp(-t^2) dt$.
Note that, as demonstrated in Fig. \ref{fig:current_fluctuation_scaled}, the scaling function is the same for both models I and II. We also perform an asymptotic analysis to obtain the behavior of $\mathcal{W}\left(y\right)$ in the two limiting cases when $y \ll 1$ and $y \gg 1$ and we find the following,
\begin{eqnarray} 
 \mathcal{W}\left(y\right) \simeq 
\left\{
\begin{array}{ll}
\vspace{0.15 cm}
  \sqrt{y/\pi},            ~~~  & {\rm for}~ y \ll 1 , \\
\vspace{0.15 cm}
 y,            ~~~  & {\rm for}~ y \gg 1. \\
\end{array}
\right.
\label{cf_scaling_function_limit}
\end{eqnarray}
 To verify our theoretical results, as in Eqs.~\eqref{bond-current-fluc-2}, \eqref{cf_scaling_function}, and \eqref{cf_scaling_function_limit},  
 we plot in Fig.~\ref{fig:current_fluctuation_scaled} the steady-state scaled bond-current flcutuations as a function of the rescaled hydrodynamic time $y=DT/L^{2}$ for models II (LLG, left-panel) and I (standard RTPs, right-panel) for various $\rho$ and $\gamma$ denoted; simulation results are represented the points. For both models, we see an excellent collapse of the simulation data points. The collapsed data (points) follow the analytically derived scaling function $\mathcal{W}(y)$ as given in Eq.~\eqref{cf_scaling_function} (line) and they exhibit sub-diffusive growth at small $y$ before crossing over to diffusive growth at large $y$, thus being consistent with the aymptotic form as in Eq.~\eqref{cf_scaling_function_limit}.

\section{Summary and concluding remarks}\label{Sec:conclusion}

In this work, we characterize steady-state current fluctuations in two minimal  models of hardcore RTPs by using a microscopic approach. Model I is the standard version of  hardcore RTPs introduced in Ref.~\cite{Soto_2014}, whereas model II is a long-ranged lattice gas (LLG) - a variant of model I  \cite{Tanmoy_condmat2022}; the latter model also has the hardcore constraint. Indeed, one great advantage of studying model II (LLG) is that, despite a lack of the knowledge of the steady-state measure having nontrivial spatial correlations, the model is amenable to analytical studies and is the primary focus of our work.
To this end, we introduce a truncation (closure) scheme in our microscopic dynamical framework to analytically calculate various dynamic quantities, which have been of interest in the past in the context of self-propelled particles. In particular, in this model we calculate exactly, within the truncation approximation, the fluctuations of time-integrated current $Q_i(T)$ across a bond $(i,i+1)$ in a time interval $T$ as well as instantaneous current $J_i(t) \equiv dQ_i(t)/dt$. We compare our theoretical results with that obtained from direct Monte Carlo simulations of models I and II and observe that the two models of RTPs share remarkably similar features.

The main results of this paper are as following. The time-integrated bond-current fluctuation exhibits subdiffusive growth at moderately large-time $1/D \ll T \ll L^{2}$, where $\langle Q_i(T)^{2} \rangle \sim T^{1/2}$, before crossing  over to a diffusive growth regime at very long time $T \gg L^{2}/D$, where $D(\rho, \gamma)$ is the density- and tumbling-rate-dependent bulk-diffusion coefficient \cite{Tanmoy_condmat2022}. 
Notably, the power-law behaviors are qualitatively similar  to that observed in symmetric exclusion processes \cite{Derrida-Sadhu-JSTAT2016}. Although the prefactors in the growth laws as a function of density and tumbling rate are model-dependent, but, they can be expressed in terms of the two macroscopic transport coefficients - the bulk-diffusion coefficient and the mobility.
Remarkably, in the limit of $L$, $T$ being large, with the dimensionless scaling variable $DT/L^{2}$  finite, we show that the growth of time-integrated bond-current fluctuation obeys a scaling law, that is presumably {\it universal}, i.e., independent of dynamical rules of the models and parameter values [see Eq. \eqref{bond-current-fluc-3}].

Furthermore, using the truncation scheme introduced in our microscopic theory, we analytically calculate in model II (LLG) the fluctuations of space-time integrated current across the entire system (equivalently, the cumulative displacement of all particles). In this way, we can characterize the current fluctuations in terms of  the collective particle mobility $\chi(\rho, \gamma) \equiv \lim_{L \rightarrow \infty} (1/2LT) \langle [\sum_{i=1}^L Q_i^2(T)] \rangle_c$ [see Eq. \ref{I_fluctuation1}], which is also equal to the scaled space-time integrated fluctuating (``noise ") current. Interestingly, in the limit of small density and strong persistence, i.e., $\rho, \gamma \rightarrow 0$ with scaling variable $\psi = \rho/\gamma \sim l_p/\gamma$ fixed, we show that, similar to the bulk-diffusion coefficient $D(\rho, \gamma)$ previously studied in Ref. \cite{Tanmoy_condmat2022}, {\it there exists a scaling regime for the mobility $\chi(\rho, \gamma)$ too.} Indeed, the scaling regime implies that in this limit, the system is governed by essentially only {\it two length scales} - the persistence length $l_p \sim \gamma^{-1}$ and the mean gap $\bar{g} \sim 1/\rho$  \cite{Tanmoy_condmat2022}. Thus our analysis highlights the role of persistence and interaction in the collective relaxation and fluctuation in the standard models of hardcore RTPs.

Moreover, our theoretical scheme readily allows us to calculate the spatial and dynamic properties of the instantaneous current in model II (LLG). Using our microscopic calculations, we derive that, at long times, two-point dynamic correlation function for instantaneous bond-current as a function of time $t$ is indeed a power-law  of the form $\sim t^{-3/2}$ and, moreover, it is negative for $t \ne 0$, with a delta-correlated part at the origin $t=0$. Similar behavior is also observed for model I (standard hardcore RTPs).
On the other hand, the spatial correlations of current in both models are found to decay exponentially $\exp(-r/\xi)$ with spatial separation $r$.
Interestingly, in the strong-persistence regime, we find that correlation length $\xi$ \textit{diverges} as the square root of persistence time $\tau_p$, i.e., $\xi \sim \sqrt{\tau_p}$, the behavior which we derive analytically for model II (LLG). This result provides a microscpoic theoretical explanations of the qualitative features of velocity correlations observed in the recent experiments and simulations \cite{Bertin_NatComm2020, Caprini_PRL_2020}.

Characterizing current fluctuations in driven many-body systems having a nontrivial spatial structure is a difficult problem in general and exact calculations of the dynamic current correlations have been done for few such systems so far \cite{Hazra_2023}. Previously, an exact calculation of the temporal growth of the time-integrated bond-current has been performed for the symmetric simple exclusion process (SSEP), which, despite having hardcore interactions, however has a product measure and, as a result, does not have spatial correlations \cite{Derrida-Sadhu-JSTAT2016}.
 However the above analysis for SSEP cannot be easily extended to systems with finite spatial correlations, which is the case here. This is because, in such a case, the equations governing the two-point correlations involve higher-order correlations, resulting in an infinite hierarchy of many-body correlations. 
 To get around the difficulty, in this paper we use a truncation scheme \cite{Anirban-PRE2023}, which, though approximate, immediately leads to a closed set of equations for density and current correlations, thus making model II (LLG) solvable.

Notably, our theory is based on two important assertions \cite{Derrida_PRL2004, MFT-RMP2015}: (i) The total instantaneous current can be decomposed into a diffusive and a fluctuating (``noise'') component, with the latter having mean zero and short-ranged (in fact, delta-correlated) temporal correlations, and (ii) the diffusive current can be written as a product of the bulk-diffusion coefficient and the gradient of mass (occupation) variable. The latter implies that the local relaxation processes are effectively ``slaved" to  coarse-grained local density, implying a ``local-equilibrium'' property of the steady state and thus a diffusive relaxation on long-time scales. Indeed our theory leads to explicit analytical calculation of various quantities, such as current fluctuations, the  mobility [see Eqs.~\eqref{bond-current-fluc-1} and \eqref{chi}] and their scaling properties [see Eqs. \eqref{bond-current-fluc-3} and \eqref{transport_scaling}] - all of which agree remarkably well with simulations [see Figs. \ref{fig:current_fluctuation_scaled} and \ref{fig:transport_coefficients_scaled}]. Furthermore, the comparison of models I and II shows that the two systems have qualitatively similar spatio-temporal behaviors.  The striking resemblance suggests that the typical models of interacting RTPs can be described by the same theoretical framework, formulated in terms of the two macroscopic transport coefficients - the bulk-diffusion coefficient and the conductivity. 
 Indeed, dynamic fluctuations in interacting SPPs have not yet been fully understood and a general theoretical formulation for dealing with such systems is still lacking.
 In this scenario, our study could provide useful insights into the large-scale structure of interacting SPPs in general.

\begin{widetext}
\section*{Appendix}

\subsection{Time evolution of two point uequal-time current-current correlation $\mathcal{C}^{Q Q}_{r}(t',t)$}\label{sec:app1}
\begin{eqnarray} 
 Q_{r}(t'+dt')Q_{0}(t) = 
\left\{
\begin{array}{ll}
\vspace{0.15 cm}
 (Q_{r}(t')  + 1)Q_{0}(t),            ~~~  & {\rm prob.}~~~ \mathcal{P}^{R}_{r} dt' , \\
\vspace{0.15 cm}
 (Q_{r}(t')  - 1)Q_{0}(t),            ~~~  & {\rm prob.}~~~ \mathcal{P}^{L}_{r} dt', \\
 \vspace{0.15 cm}
 Q_{r}(t')Q_{0}(t),                ~~~  & {\rm prob.}~~~~  1 - (\mathcal{P}^{R}_{r} + \mathcal{P}^{L}_{r})dt'. \\
\end{array}
 \right.
\label{Q1_Q2_update_eq_unequal}
\end{eqnarray}
Using the update rules in Eq.~\eqref{Q1_Q2_update_eq_unequal} and substituting the expressions of $\mathcal{P}^{R}_{r}$ and $\mathcal{P}^{L}_{r}$, as shown in Eqs.~\eqref{pr} and \eqref{pl}, respectively, the corresponding time-evolution equation can be written as,
\begin{eqnarray}
\frac{d}{dt'}\left\langle Q_{r}(t')Q_{0}(t) \right\rangle=\frac{1}{2}\sum_{l=1}^{\infty}\phi(l)\left[\Big\{ \left\langle \hat{\mathcal{U}}^{(l)}_{r+l}(t')Q_{0}(t) \right\rangle - \left\langle \hat{\mathcal{U}}^{(l)}_{r}(t')Q_{0}(t) \right\rangle \Big\} + \sum_{g=1}^{l-1} \Big\{\left\langle \hat{\mathcal{V}}^{(g+2)}_{r+g+1}(t')Q_{0}(t) \right\rangle - \left\langle \hat{\mathcal{V}}^{(g+2)}_{r+1}(t')Q_{0}(t) \right\rangle \Big\} \right],\nonumber \\
\end{eqnarray}
Finally using the definition of $\mathcal{C}^{Q Q}_{r}(t',t)$ as provided in Eq.~\eqref{current_correlation_defn}, we immediately obtain
\begin{eqnarray}\label{time-evolution-Q-Q-general_app}
\frac{d}{dt'}\mathcal{C}^{QQ}_{r}(t',t)  &=& \frac{1}{2}\sum_{l=1}^{\infty}\phi(l)\Big[ \Big\{ \mathcal{C}^{\mathcal{U}^{(l)}Q}_{r+l}(t',t) - \mathcal{C}^{\mathcal{U}^{(l)}Q}_{r}(t',t) \Big\} + \sum_{g=1}^{l-1} \Big\{\mathcal{C}^{\mathcal{V}^{(g+2)}Q}_{r+g+1}(t',t) - \mathcal{C}^{\mathcal{V}^{(g+2)}Q}_{r+1}(t',t) \Big\}\Big],  \\ &=& \left \langle J^{(D)}_{r}(t')Q_{0}(t)\right \rangle_{c}.
\label{time-evolution-Q-Q-general1_app}
\end{eqnarray}
Here we have used the expression for $J^{(D)}_{r}(t')$ as given in Eq.~\eqref{diffusive_current}. Note that the above two equations are expressed in the main text as Eqs.~\eqref{time-evolution-Q-Q-general} and \eqref{time-evolution-Q-Q-general1}.
\vspace{0.5 cm}
\subsection{Time evolution of the two point uequal-time density-current correlation $\mathcal{C}^{\eta Q}_{r}(t',t)$}\label{sec:app2}
In this section, we derive the time-evolution equation for the two point uequal-time density-current correlation $\mathcal{C}^{\eta Q}_{r}(t',t)$ as shown in Eq.~\eqref{time-evolution-eta-Q-diff_time} in the main text. In order to do so, we first derive the time-evolution equation of the local density $\rho_{r}(t)$, which is defined as the average occupancy at site $r$ and time $t$, i.e., $\rho_{r}(t)=\langle \eta_{r}(t)\rangle$.\\
Recall, the average instantaneous bond-current $\left \langle J_{r}(t) \right \rangle$  across the bond $(r, r+1)$ at time $t$ (as given by Eq.~\eqref{time-derivative-int-current_2} in the main text) is given by,
\begin{eqnarray}\label{time-derivative-int-current_2_app}
\left \langle J_{r}(t) \right \rangle = \frac{1}{2}\sum_{l=1}^{\infty}\phi(l)\left[\sum_{g=1}^{l-1}\left( \mathcal{V}_{r+g+1}^{(g+2)} - \mathcal{V}_{r+1}^{(g+2)} \right) + \left( \mathcal{U}_{r+l}^{(l)} -\mathcal{U}_r^{(l)} \right) \right].
\end{eqnarray}
Since the total number of particle is a conserved quantity, the corresponding local density is a slow variable and its time-evolution must be related to  $\left \langle J_{r}(t) \right \rangle$ via the following continuity equation:
\begin{eqnarray}\label{density_evolution_1}
\frac{d\rho_{r}(t)}{dt}=\left \langle J_{r-1}(t) \right \rangle-\left \langle J_{r}(t) \right \rangle.
\end{eqnarray}
Using Eq.~\eqref{time-derivative-int-current_2_app} in Eq.~\eqref{density_evolution_1}, it is now straighforward to obtain the corresponding time-evolution equation in the form of the following balance equation:
\begin{eqnarray}\label{density_evolution_2}
\frac{d\rho_{r}(t)}{dt}=\left \langle \mathcal{P}^{+}_{r}(t) \right \rangle-\left \langle \mathcal{P}^{-}_{r}(t) \right \rangle,
\end{eqnarray}
where the gain and loss terms are respectively given by,
\begin{eqnarray}\label{gain-loss_terms}
\mathcal{P}^{+}_{r}(t) &=& \frac{1}{2}\sum_{l=1}^{\infty}\phi(l)\left[\sum_{g=1}^{l-1}\Big\{\mathcal{V}_{r+g}^{(g+2)} + \mathcal{V}_{r+1}^{(g+2)} \Big\} + \Big\{\left( \mathcal{U}_{r+l-1}^{(l)} -\mathcal{U}_{r+l}^{(l+1)} \right) + \left( \mathcal{U}_{r}^{(l)} -\mathcal{U}_{r}^{(l+1)} \right) \Big\} \right], \\
\mathcal{P}^{-}_{r}(t) &=& \frac{1}{2}\sum_{l=1}^{\infty}\phi(l)\left[\sum_{g=1}^{l-1}\Big\{\mathcal{V}_{r+g+1}^{(g+2)} + \mathcal{V}_{r}^{(g+2)} \Big\} + \Big\{\left( \mathcal{U}_{r+l}^{(l)} -\mathcal{U}_{r+l}^{(l+1)} \right) + \left( \mathcal{U}_{r-1}^{(l)} -\mathcal{U}_{r}^{(l+1)} \right) \Big\} \right].
\end{eqnarray}
We now use the expression of the local diffusive current operator, as shown in Eq.~\eqref{diffusive_current} in the main text, to arrive at the following identity:
\begin{eqnarray}
\mathcal{P}^{+}_{r}(t)-\mathcal{P}^{-}_{r}(t)&=&J^{(D)}_{r-1}(t)-J^{(D)}_{r}(t),\\ &\simeq& D(\rho, \gamma) \left[\eta_{r+1}(t)+\eta_{r-1}(t)-2\eta_{r}(t)\right],
\label{identity_2}
\end{eqnarray} 
where Eq.~\eqref{identity_2} is a direct consequence of the proposed closure approximation, as shown in Eq.~\eqref{closure_approximation} in the main text. Finally using Eqs.~\eqref{identity_2} and \eqref{density_evolution_2} together, we arrive at the desired time-evolution equation,
\begin{eqnarray}\label{density_evolution_3}
\frac{d\rho_{r}(t)}{dt} \simeq D(\rho, \gamma) \left[\rho_{r+1}(t)+\rho_{r-1}(t)-2\rho_{r}(t)\right]=D(\rho, \gamma) \Delta^{2}_{r}\rho_{r}(t),
\end{eqnarray}
where $\Delta^{2}_{r}$ is the discrete laplacian operator. We will now deduce the desired time-evolution of $\mathcal{C}^{\eta Q}_{r}(t',t)$ by writing the following update rules:
\begin{eqnarray} 
 \eta_{r}(t'+dt')Q_{0}(t) = 
\left\{
\begin{array}{ll}
 \vspace{0.15 cm}
 1\times Q_{0}(t),            ~~~  & {\rm prob.}~~~ \mathcal{P}_{r}^{+}(t') dt',
 \\
 \vspace{0.15 cm}
 0\times Q_{0}(t),            ~~~  & {\rm prob.}~~~ \mathcal{P}_{r}^{-}(t') dt', \\
\vspace{0.15 cm}
 \eta_{r}(t)Q_{0}(t),                ~~~  & {\rm prob.}~~~~  1 - \Sigma dt, \\
\end{array}
 \right.
\label{eta1_Q2_update_unequal_time}
\end{eqnarray}
Using the update equation, as shown above in Eq.~\eqref{eta1_Q2_update_unequal_time}, we write down the corresponding time-evolution equation as,
\begin{eqnarray}
\frac{d}{dt'}\left\langle \eta_{r}(t') Q_{0}(t) \right\rangle &=& \left\langle \left( \mathcal{P}^{+}_{r}(t')-\mathcal{P}^{-}_{r}(t') \right)Q_{0}(t) \right\rangle. \\ &\simeq& D(\rho,\gamma) \Delta^{2}_{r}\left\langle \eta_{r}(t') Q_{0}(t) \right\rangle,
\end{eqnarray}
where in the last equation, we have used the identity displayed in Eq.~\eqref{identity_2}. Now, by using the definition of $\mathcal{C}^{\eta Q}_{r}(t',t)=\langle \eta_{r}(t') Q_{0}(t) \rangle - \langle \eta_{r}(t') \rangle Q_{0}(t) \rangle$, we directly obtain
\begin{eqnarray}\label{eta_Q_evolution_unequal_time_app}
\frac{d}{dt'}\mathcal{C}^{\eta Q}_{r}(t',t) \simeq D(\rho,\gamma) \Delta^{2}_{r}\mathcal{C}^{\eta Q}_{r}(t',t).
\end{eqnarray}
Note that Eq.~\eqref{eta_Q_evolution_unequal_time_app} is the desired time-evolution equation which we have used in Eq.~\eqref{time-evolution-eta-Q-diff_time} in the main text.
\subsection{Time evolution of the equal time density-current correlation $\mathcal{C}^{\eta Q}_{r}(t,t)$}\label{sec:app3}
Here we will derive the time-evolution equation for the equal time density-current correlation $\mathcal{C}^{\eta Q}_{r}(t,t)$, which is presented in Eq.~\eqref{eta1_Q2_evolution_compact} of the main text. We write down all of the possible ways that the product $\eta_{r}Q_{0}$ can change in the infinitesimal time interval $[t, t + dt]$, as given by
\begin{eqnarray} 
 \eta_{r}(t+dt)Q_{0}(t+dt) = 
\left\{
\begin{array}{ll}
\vspace{0.15 cm}
 1\times(Q_{0}(t)  + 1),            ~~~  & {\rm prob.}~~~ \mathcal{P}_r^{3}(t) dt , \\
\vspace{0.15 cm}
 1\times (Q_{0}(t)-1),            ~~~  & {\rm prob.}~~~ \mathcal{P}_r^{4}(t) dt, \\
 \vspace{0.15 cm}
 0\times(Q_{0}(t)  + 1),            ~~~  & {\rm prob.}~~~ \mathcal{P}_r^{5}(t) dt , \\
\vspace{0.15 cm}
 0\times (Q_{0}(t)-1),            ~~~  & {\rm prob.}~~~ \mathcal{P}_r^{6}(t) dt, \\
 \vspace{0.15 cm}
  \eta_{r}(t)(Q_{0}(t)  + 1),            ~~~  & {\rm prob.}~~~ [\mathcal{P}_{0}^R(t) - \mathcal{P}_r^{3}(t)-\mathcal{P}_r^{5}(t)] dt, \\
 \vspace{0.15 cm}
 \eta_{r}(t)(Q_{0}(t)  - 1),            ~~~  & {\rm prob.}~~~ [\mathcal{P}_{0}^L(t) - \mathcal{P}_r^{4}(t) -\mathcal{P}_r^{6}(t)] dt,
 \\
 \vspace{0.15 cm}
 1\times Q_{0}(t),            ~~~  & {\rm prob.}~~~ [\mathcal{P}^{+}_r(t) - \mathcal{P}_r^{3}(t)-\mathcal{P}_r^{4}(t)] dt,
 \\
 \vspace{0.15 cm}
 0\times Q_{0}(t),            ~~~  & {\rm prob.}~~~ [\mathcal{P}^{-}_r(t) - \mathcal{P}_r^{5}(t)-\mathcal{P}_r^{6}(t)] dt, \\
\vspace{0.15 cm}
 \eta_{r}(t)Q_{0}(t),                ~~~  & {\rm prob.}~~~~  1 - \hat{\Sigma}(t) dt, \\
\end{array}
 \right.
\label{eta1_Q2_update_eq}
\end{eqnarray}
where $\hat{\Sigma}(t) dt$ is the sum of probability operators corresponding to the all possible ways in which the product $\eta_{r}(t)Q_{0}(t)$ can change in the infinitesimal time interval $dt$ and is given by
\begin{eqnarray}\label{sigma_eta_q}
\hat{\Sigma}(t) = \mathcal{P}^{+}_r(t) + \mathcal{P}^{-}_r(t) + \mathcal{P}_{0}^R(t) + \mathcal{P}_{0}^L(t) + \mathcal{P}_r^{3}(t) + \mathcal{P}_r^{4}(t) + \mathcal{P}_r^{5}(t) + \mathcal{P}_r^{6}(t),
\end{eqnarray}
and the operators $\mathcal{P}_r^{3}(t)$, $\mathcal{P}_r^{4}(t)$, $\mathcal{P}_r^{5}(t)$ and $\mathcal{P}_r^{6}(t)$ are defined as,
\begin{eqnarray}\label{operators_2}
\mathcal{P}_r^{3}(t)&=&\frac{1}{2}\sum_{l=1}^{\infty} \phi(l) \left[ \left(\mathcal{U}_{r}^{(l)} - \mathcal{U}_{r}^{(l+1)}\right)\sum_{k=1}^{l}\delta_{r,k} + \sum_{g=1}^{l-1}\mathcal{V}_{r+1}^{(g+2)}\sum_{k=1}^{g}\delta_{r,k}\right], \\
\mathcal{P}_r^{4}(t)&=&\frac{1}{2}\sum_{l=1}^{\infty} \phi(l) \left[ \left(\mathcal{U}_{r+l-1}^{(l)} - \mathcal{U}_{r+l}^{(l+1)}\right)\sum_{k=1}^{l}\delta_{r,-k+1} + \sum_{g=1}^{l-1}\mathcal{V}_{r+g}^{(g+2)}\sum_{k=1}^{g}\delta_{r,-k+1}\right],\\
\mathcal{P}_r^{5}(t)&=&\frac{1}{2}\sum_{l=1}^{\infty} \phi(l) \left[ \left(\mathcal{U}_{r+l}^{(l)} - \mathcal{U}_{r+l}^{(l+1)}\right)\sum_{k=1}^{l}\delta_{r,-k+1} + \sum_{g=1}^{l-1}\mathcal{V}_{r+g+1}^{(g+2)}\sum_{k=1}^{g}\delta_{r,-k+1}\right],\\
\mathcal{P}_r^{6}(t)&=&\frac{1}{2}\sum_{l=1}^{\infty} \phi(l) \left[ \left(\mathcal{U}_{r-1}^{(l)} - \mathcal{U}_{r}^{(l+1)}\right)\sum_{k=1}^{l}\delta_{r,k} + \sum_{g=1}^{l-1}\mathcal{V}_{r}^{(g+2)}\sum_{k=1}^{g}\delta_{r,k}\right].
\end{eqnarray}
Using the above update rules, shown in Eq.~\eqref{eta1_Q2_update_eq}, the time-evolution of the quantity $\langle \eta_{r}(t) Q_{0}(t) \rangle$ is given by,
\begin{eqnarray}\label{eta1_Q2_evolution_equation_app}
\frac{d}{dt}\left\langle \eta_{r}(t) Q_{0}(t) \right\rangle &=& \left[\left\langle\mathcal{P}_r^{3}(t) \right\rangle - \left\langle \mathcal{P}_r^{4}(t) \right\rangle -\left\langle \mathcal{P}_r^{5}(t) \right\rangle + \left\langle \mathcal{P}_r^{6}(t) \right\rangle\right] + \left\langle\left[\mathcal{P}^{+}_{r}(t)-\mathcal{P}^{-}_{r}(t)\right]Q_{0}(t)\right\rangle + \left\langle \eta_{r}(t) \left[\mathcal{P}_{0}^R(t)-\mathcal{P}_{0}^L(t)\right]\right\rangle.\nonumber \\
\end{eqnarray}
At the steady-state, we can disregard the spatial dependence in the average quantities $\langle \mathcal{U}^{(l)} \rangle$ and $\langle \mathcal{V}^{(g+2)} \rangle$, which essentially leads to the simplification of the first term in Eq.~\eqref{eta1_Q2_evolution_equation_app} in the following manner:
\begin{eqnarray}
\left\langle\mathcal{P}_r^{3}(t) \right\rangle - \left\langle \mathcal{P}_r^{4}(t) \right\rangle -\left\langle \mathcal{P}_r^{5}(t) \right\rangle + \left\langle \mathcal{P}_r^{6}(t) \right\rangle =\sum_{l=1}^{\infty} \phi(l) \left[ \left(\mathcal{U}^{(l)} - \mathcal{U}^{(l+1)}\right)\sum_{k=1}^{l}(\delta_{r,k}-\delta_{r,-k+1}) + \sum_{g=1}^{l-1}\mathcal{V}^{(g+2)}\sum_{k=1}^{g}(\delta_{r,k}-\delta_{r,-k+1})\right].\nonumber \\
\end{eqnarray}
Moreover, using the identity shown in Eq.~\eqref{identity_2}, the second term in Eq.~\eqref{eta1_Q2_evolution_equation_app} can be transformed into
\begin{eqnarray}
\left\langle\left[\mathcal{P}^{+}_{r}(t)-\mathcal{P}^{-}_{r}(t)\right]Q_{0}(t)\right\rangle \simeq D(\rho, \gamma) \Delta^{2}_{r} \left\langle \eta_{r}(t) Q_{0}(t) \right\rangle. 
\end{eqnarray}
Furthermore, using the following relation: $\mathcal{P}_{0}^R(t)-\mathcal{P}_{0}^L(t) = J^{(D)}_{0}(t) \simeq D(\rho, \gamma) (\eta_{0}-\eta_{1})$, we rewrite the third term in Eq.~\eqref{eta1_Q2_evolution_equation_app} in the following way:
\begin{eqnarray}
\left\langle \eta_{r}(t) \left[\mathcal{P}_{0}^R(t)-\mathcal{P}_{0}^L(t)\right]\right\rangle \simeq D(\rho,\gamma) \left[ \left\langle \eta_{r}(t) \eta_{0}(t) \right\rangle - \left\langle \eta_{r}(t) \eta_{1}(t) \right\rangle \right]=D(\rho,\gamma) \Delta_{r}\left\langle \eta_{r}(t) \eta_{0}(t) \right\rangle.
\end{eqnarray}
Finally using the last three transformations, the time-evolution equation of $\mathcal{C}^{\eta Q}_{r}(t,t)$ can be written as the following inhomogeneous differential equation:
\begin{eqnarray}\label{eta_Q_same_time_evolution_app}
\frac{d}{dt}\mathcal{C}^{\eta Q}_{r}(t,t)\simeq D(\rho,\gamma)\Delta^{2}_{r}\mathcal{C}^{\eta Q}_{r}(t,t) + \mathcal{S}^{\eta Q}_{r}(t),
\end{eqnarray}
where the source term is given by
\begin{eqnarray}\label{eta_Q_source_real_app}
\mathcal{S}^{\eta Q}_{r}(t)= \sum_{l=1}^{\infty} \phi(l) \left[ \left(\mathcal{U}^{(l)} - \mathcal{U}^{(l+1)}\right)\sum_{k=1}^{l}(\delta_{r,k}-\delta_{r,-k+1}) + \sum_{g=1}^{l-1}\mathcal{V}^{(g+2)}\sum_{k=1}^{g}(\delta_{r,k}-\delta_{r,-k+1})\right] + D(\rho,\gamma) \Delta_{r}\mathcal{C}^{\eta \eta}_{r}(t,t).\nonumber \\
\end{eqnarray}
Eq.~\eqref{eta_Q_same_time_evolution_app} describes the time-evolution of $\mathcal{C}^{\eta Q}_{r}(t,t)$ in the real space, which in the Fourier space is simply transformed into the following equation:
\begin{eqnarray}\label{eta1_Q2_evolution_compact_app}
\left(\frac{d}{dt} + D(\rho,\gamma) \lambda_{q} \right)\mathcal{\tilde{C}}^{\eta Q}_{q}(t,t)=\mathcal{\tilde{S}}^{\eta Q}_{q}(t).
\end{eqnarray}
Note that, the above equation appears in Eq.~\eqref{eta1_Q2_evolution_compact} in the main text. Here $-\lambda_{q}$ is the eigen-value of the discrete laplacian operator, which is given by,
\begin{eqnarray}
\lambda_q =2\left(1- \cos(q) \right),
\end{eqnarray}
and $\mathcal{\tilde{S}}^{\eta Q}_{q}(t)$ is the source term in the Fourier space which is trivially obtained to be,
\begin{eqnarray}\label{source-eta-Q_app}
\mathcal{\tilde{S}}^{\eta Q}_{q}(t) &=& \frac{1}{(1-e^{-iq})}\left[D(\rho,\gamma) \lambda_{q} \mathcal{\tilde{C}}^{\eta \eta}_{q}(t,t) -\sum_{l=1}^{\infty} \phi(l) \Big\{ \left(\mathcal{U}^{(l)} - \mathcal{U}^{(l+1)}\right)\lambda_{lq} + \sum_{g=1}^{l-1}\mathcal{V}^{(g+2)}\lambda_{gq}\Big\}\right]. 
\end{eqnarray}
Finally using the following identities that relates the correlators $\mathcal{U}^{(l)}$, $\mathcal{V}^{(g+2)}$ with the gap-distrbution function $P(g)$ as given by
\begin{eqnarray}\label{identity}
\mathcal{U}^{(l)}(t)&=& \rho \sum_{g=l-1}^{\infty} (g-l+1) P(g,t),\nonumber \\ 
\mathcal{V}^{(g+2)}(t)&=&\rho P(g,t),
\end{eqnarray}
we obtain simpler structure of $\mathcal{\tilde{S}}^{\eta Q}_{q}(t)$, which is given by
\begin{eqnarray}\label{source-eta-Q_app}
\mathcal{\tilde{S}}^{\eta Q}_{q}(t) &=& \frac{1}{(1-e^{-iq})}\left[D(\rho,\gamma) \lambda_{q} \mathcal{\tilde{C}}^{\eta \eta}_{q}(t,t) -f_{q}(t)\right], 
\end{eqnarray}
where $f_{q}(t)$ is given by
\begin{eqnarray}\label{f_q_app}
f_{q}(t)=\rho \sum_{l=1}^{\infty}\phi(l)\left[\sum_{g=1}^{l-1}  \lambda_{gq}P(g, t) +  \lambda_{lq}\sum_{g=l}^{\infty}P(g, t)\right].
\end{eqnarray}
The last two equations compositely express the source term corresponding to the equal time density-current correlation function and they appear in the main text in Eqs.~\eqref{source-eta-Q} and \eqref{f_{q}}, respectively.
\subsection{Time evolution of equal time density-density correlation $\mathcal{C}^{\eta \eta}_{r}(t,t)$}\label{sec:app4}
\begin{eqnarray} 
 \eta_{r}(t+dt)\eta_{0}(t+dt) = 
\left\{
\begin{array}{ll}
\vspace{0.15 cm}
 1 \times 1,            ~~~  & {\rm prob.}~~~ \mathcal{P}_r^{7}(t) dt , \\
\vspace{0.15 cm}
 0 \times 0,            ~~~  & {\rm prob.}~~~ \mathcal{P}_r^{8}(t) dt, \\
 \vspace{0.15 cm}
 1\times 0,            ~~~  & {\rm prob.}~~~ \mathcal{P}_r^{9}(t) dt , \\
\vspace{0.15 cm}
 0\times 1,            ~~~  & {\rm prob.}~~~ \mathcal{P}_r^{10}(t) dt, \\
 \vspace{0.15 cm}
  1 \times \eta_{0}(t),            ~~~  & {\rm prob.}~~~ [\mathcal{P}_{r}^+(t) - \mathcal{P}_r^{7}(t)-\mathcal{P}_r^{9}(t)] dt, \\
 \vspace{0.15 cm}
 0 \times \eta_{0}(t),            ~~~  & {\rm prob.}~~~ [\mathcal{P}_{r}^-(t) - \mathcal{P}_r^{8}(t) -\mathcal{P}_r^{10}(t)] dt,
 \\
 \vspace{0.15 cm}
 \eta_{r}(t) \times 1,            ~~~  & {\rm prob.}~~~ [\mathcal{P}^{+}_0(t) - \mathcal{P}_r^{7}(t)-\mathcal{P}_r^{10}(t)] dt,
 \\
 \vspace{0.15 cm}
 \eta_{r}(t) \times 0,            ~~~  & {\rm prob.}~~~ [\mathcal{P}^{-}_0(t) - \mathcal{P}_r^{8}(t)-\mathcal{P}_r^{9}(t)] dt, \\
\vspace{0.15 cm}
 \eta_{r}(t)\eta_{0}(t),                ~~~  & {\rm prob.}~~~~  1 - \hat{\Sigma}(t) dt, \\
\end{array}
 \right.
\label{eta1_eta2_update_eq}
\end{eqnarray}
where $\hat{\Sigma}(t) dt$ corresponds to the total probability with which the product of occupations at sites $r$ and $0$ changes in the infinitesimal time interval $dt$ with
\begin{eqnarray}
\hat{\Sigma}(t)= \mathcal{P}_{r}^+(t) + \mathcal{P}_{r}^-(t) + \mathcal{P}_{0}^+(t) + \mathcal{P}_{0}^-(t)-\mathcal{P}_r^{7}(t)-\mathcal{P}_r^{8}(t)-\mathcal{P}_r^{9}(t)-\mathcal{P}_r^{10}(t),
\end{eqnarray}
and the operators $\mathcal{P}_r^{7}(t)$, $\mathcal{P}_r^{8}(t)$, $\mathcal{P}_r^{9}(t)$ and $\mathcal{P}_r^{10}(t)$ are defined as,
\begin{eqnarray}
\mathcal{P}_r^{7}(t) &=& \frac{1}{2}\sum_{l=1}^{\infty}\phi(l)\left[\sum_{g=1}^{l-1}\left(\mathcal{V}^{(g+2)}_{r+1} + \mathcal{V}^{(g+2)}_{r+g} \right)\delta_{r,0} + \Big\{\left(\mathcal{U}^{(l)}_{r} - \mathcal{U}^{(l+1)}_{r} \right) + \left(\mathcal{U}^{(l)}_{r+l-1} - \mathcal{U}^{(l+1)}_{r+l} \right) \Big\} \delta_{r,0}\right], \\
\mathcal{P}_r^{8}(t) &=& \frac{1}{2}\sum_{l=1}^{\infty}\phi(l)\left[\sum_{g=1}^{l-1}\left(\mathcal{V}^{(g+2)}_{r} + \mathcal{V}^{(g+2)}_{r+g+1} \right)\delta_{r,0} + \Big\{\left(\mathcal{U}^{(l)}_{r+l} - \mathcal{U}^{(l+1)}_{r+l} \right) + \left(\mathcal{U}^{(l)}_{r-1} - \mathcal{U}^{(l+1)}_{r} \right) \Big\} \delta_{r,0}\right],\\
\mathcal{P}_r^{9}(t) &=& \frac{1}{2}\sum_{l=1}^{\infty}\phi(l)\left[\sum_{g=1}^{l-1}\left(\mathcal{V}^{(g+2)}_{g+1}\delta_{r,g} + \mathcal{V}^{(g+2)}_{0} \delta_{r,-g}\right) + \Big\{\left(\mathcal{U}^{(l)}_{l} - \mathcal{U}^{(l+1)}_{l} \right)\delta_{r,l} + \left(\mathcal{U}^{(l)}_{-1} - \mathcal{U}^{(l+1)}_{0} \right) \delta_{r,-l} \Big\}\right],\\
\mathcal{P}_r^{10}(t) &=& \frac{1}{2}\sum_{l=1}^{\infty}\phi(l)\left[\sum_{g=1}^{l-1}\left(\mathcal{V}^{(g+2)}_{r+g+1}\delta_{r,-g} + \mathcal{V}^{(g+2)}_{r} \delta_{r,g}\right) + \Big\{\left(\mathcal{U}^{(l)}_{r+l} - \mathcal{U}^{(l+1)}_{r+l} \right)\delta_{r,-l} + \left(\mathcal{U}^{(l)}_{r-1} - \mathcal{U}^{(l+1)}_{r} \right) \delta_{r,l} \Big\}\right].
\end{eqnarray} 
Finally using the above update rules in Eq.~\eqref{eta1_eta2_update_eq}, we can straighforwardly write down the corresponding time-evolution equiation, which is given by,
\begin{eqnarray}\label{density_density_update_eq_time_app_1}
\frac{d}{dt}\left\langle\eta_{r}(t)\eta_{0}(t)\right\rangle = \left(\left\langle \mathcal{P}_r^{7}(t) \right\rangle + \left\langle \mathcal{P}_r^{8}(t) \right\rangle - \left\langle \mathcal{P}_r^{9}(t) \right\rangle - \left\langle \mathcal{P}_r^{10}(t) \right\rangle \right) + \left\langle \eta_{r}(t) \left(\mathcal{P}^{+}_0(t) - \mathcal{P}^{-}_0(t)\right)\right\rangle + \left\langle  \left(\mathcal{P}^{+}_r(t) - \mathcal{P}^{-}_r(t)\right)\eta_{0}(t)\right\rangle.\nonumber \\
\end{eqnarray} 
Using the concept of spatial homogeneity at the steady-state, we can now ignore the spatial dependence in the averages $\left\langle \mathcal{U}^{(l)} \right\rangle$ and $\left\langle \mathcal{V}^{(g+2)} \right\rangle$, which leads to the following simplification of the first term in Eq.~\eqref{density_density_update_eq_time_app_1}:
\begin{eqnarray}
\mathcal{S}^{\eta \eta}_{r}(t) &=& \left\langle \mathcal{P}_r^{7}(t) \right\rangle + \left\langle \mathcal{P}_r^{8}(t) \right\rangle - \left\langle \mathcal{P}_r^{9}(t) \right\rangle - \left\langle \mathcal{P}_r^{10}(t) \right\rangle,\nonumber \\
&=& \sum_{l=1}^{\infty}\phi(l)\left[\sum_{g=1}^{l-1} \mathcal{V}^{(g+2)}(t)\left( 2\delta_{r,0} - \delta_{r,g} - \delta_{r,-g}\right) + \left( \mathcal{U}^{(l)}(t)- \mathcal{U}^{(l+1)}(t) \right)\left( 2\delta_{r,0} - \delta_{r,l} - \delta_{r,-l}\right)  \right],\nonumber \\
&=& \rho \sum_{l=1}^{\infty}\phi(l)\left[\sum_{g=1}^{l-1} P(g,t)\left( 2\delta_{r,0} - \delta_{r,g} - \delta_{r,-g}\right) +\sum_{g=l}^{\infty}P(g,t) \left( 2\delta_{r,0} - \delta_{r,l} - \delta_{r,-l}\right)  \right].
\label{source_eta_eta_app}
\end{eqnarray}
Note that, in the last line we have used the identities in Eq.~\eqref{identity} to replace the correlators $\mathcal{V}^{(g+2)}(t)$ and $\mathcal{U}^{(l)}(t)$ in terms of the gap-distribution function $P(g,t)$. Furthermore, using the second identity, shown in Eq.~\eqref{identity_2}, and the definition of $\mathcal{C}^{\eta \eta}_{r}(t,t)$, we can write down the corresponding time-evolution equation in real space in the following manner: 
\begin{eqnarray}\label{density_density_update_eq_time_app_2}
\frac{d}{dt}\mathcal{C}^{\eta \eta}_{r}(t,t) = 2D(\rho,\gamma)\Delta_{r}^{2}\mathcal{C}^{\eta \eta}_{r}(t,t) + \mathcal{S}^{\eta \eta}_{r}(t).
\end{eqnarray} 
In fact, using Eq.~\eqref{Fourier_transform} in the above equation, we immediately obtain the corresponding evolution equation in the Fourier space, which is given by,
\begin{eqnarray}\label{density_density_update_eq_time_app_3}
\frac{d}{dt}\mathcal{\tilde{C}}^{\eta \eta}_{q}(t,t) + 2D(\rho,\gamma)\lambda_{q}\mathcal{\tilde{C}}^{\eta \eta}_{q}(t,t) = \mathcal{\tilde{S}}^{\eta \eta}_{q}(t),
\end{eqnarray}
where $\mathcal{\tilde{S}}^{\eta \eta}_{q}(t)$ is the corresponding source-term in the Fourier space, which is simply obtained by using Eq.~\eqref{Fourier_transform} in Eq.~\eqref{source_eta_eta_app} and is given by,
\begin{eqnarray}\label{source_eta_eta_app_1}
\mathcal{\tilde{S}}^{\eta \eta}_{q}(t)=\rho \sum_{l=1}^{\infty}\phi(l)\left[\sum_{g=1}^{l-1}  \lambda_{gq}P(g, t) +  \lambda_{lq}\sum_{g=l}^{\infty}P(g, t)\right].
\end{eqnarray}
It is worth noting that Eqs.~\eqref{density_density_update_eq_time_app_3} is the desired equation used in the main text as Eq.~\eqref{eta1_eta2_evolution_compact}.
\subsection{Time evolution of equal time current-current correlation $\mathcal{C}^{Q Q}_{r}(t,t)$}\label{sec:app5}
Here we provide the derivation details of Eq.~\eqref{Q1_Q2_evolution_compact} in the main text, which describes the time-evolution of $\mathcal{C}^{Q Q}_{r}(t,t)$. We begin with the update rules corresponding to $Q_{r}(t)Q_{0}(t)$, as written below,
\begin{eqnarray} 
 Q_{r}(t+dt)Q_{0}(t+dt) = 
\left\{
\begin{array}{ll}
\vspace{0.15 cm}
 (Q_{r}(t)  + 1)(Q_{0}(t)  + 1),            ~~~  & {\rm prob.}~~~ \mathcal{P}_r^{1}(t) dt , \\
\vspace{0.15 cm}
 (Q_{r}(t)  + 1)Q_{0}(t),            ~~~  & {\rm prob.}~~~ [\mathcal{P}_r^{R}(t) - \mathcal{P}_r^{1}(t)] dt, \\
 \vspace{0.15 cm}
 Q_{r}(t)(Q_{0}(t)  + 1),            ~~~  & {\rm prob.}~~~ [\mathcal{P}_0^{R}(t) - \mathcal{P}_r^{1}(t)] dt,
 \\
 \vspace{0.15 cm}
 (Q_{r}(t)  - 1)(Q_{0}(t)  - 1),            ~~~  & {\rm prob.}~~~ \mathcal{P}_r^{2}(t) dt , \\
\vspace{0.15 cm}
 (Q_{r}(t)  - 1)Q_{0}(t),            ~~~  & {\rm prob.}~~~ [\mathcal{P}_r^{L}(t) - \mathcal{P}_r^{2}] dt, \\
 \vspace{0.15 cm}
 Q_{r}(t)(Q_{0}(t)  - 1),            ~~~  & {\rm prob.}~~~ [\mathcal{P}_0^{L}(t) - \mathcal{P}_r^{2}(t)] dt,
 \\
 \vspace{0.15 cm}
 Q_{r}(t)Q_{0}(t),                ~~~  & {\rm prob.}~~~~  1 - \hat{\Sigma}(t) dt, \\
\end{array}
 \right.
\label{Q1_Q2_update_eq}
\end{eqnarray}
where $\hat{\Sigma}(t) dt$ corresponds to the total probability with which the product of currents across bonds $(r,r+1)$ and $(0,1)$ changes in the infinitesimal time interval $dt$ with

\begin{eqnarray}\label{sigma_Q1Q2}
\hat{\Sigma}=\mathcal{P}_r^{R}(t) + \mathcal{P}_0^{R}(t)) +\mathcal{P}_r^{L}(t) + \mathcal{P}_0^{L}(t)) - \mathcal{P}_r^{1}(t) - \mathcal{P}_r^{2}(t),
\end{eqnarray}
and the operators $\mathcal{P}_r^{1}$ and $\mathcal{P}_r^{2}$ are defined as,
\begin{eqnarray}\label{P_1}
\mathcal{P}_r^{1}= \frac{1}{2}\hspace{-0.1 cm}\sum_{l=1}^{\infty} \phi(l)\Bigg\{\sum_{k=1}^{l} \left(\mathcal{U}_{r+k}^{(l)} - \mathcal{U}_{r+k}^{(l+1)}\right) \Theta(k+r)\Theta(l-r-k+1) + \sum_{g=1}^{l-1}\sum_{k=1}^{g} \mathcal{V}_{r+k+1}^{(g+2)} \Theta(k+r)\Theta(g-r-k+1)\Bigg\}, \nonumber
\end{eqnarray}
\begin{eqnarray}\label{P_2}
\mathcal{P}_r^{2}= \frac{1}{2}\hspace{-0.1 cm}\sum_{l=1}^{\infty} \phi(l)\Bigg\{\sum_{k=1}^{l} \left(\mathcal{U}_{r+k-1}^{(l)} - \mathcal{U}_{r+k}^{(l+1)}\right) \Theta(k+r)\Theta(l-r-k+1) + \sum_{g=1}^{l-1}\sum_{k=1}^{g} \mathcal{V}_{r+k}^{(g+2)} \Theta(k+r)\Theta(g-r-k+1)\Bigg\}. \nonumber \\
\end{eqnarray}
Here $\Theta(r)$ is the Heaviside theta function, which is defined as
\begin{eqnarray} 
 \Theta(r) = 
\left\{
\begin{array}{ll}
\vspace{0.15 cm}
  1,            ~~~  & {\rm for}~ r>0 , \\
\vspace{0.15 cm}
 0,            ~~~  & \text{otherwise}. \\
\end{array}
\right.
\label{theta_function}
\end{eqnarray}
Using the update rules shown in Eq.~\eqref{Q1_Q2_update_eq}, we write the evolution of two point equal-time current-current correlation as,
\begin{eqnarray}
\frac{d}{dt}\left\langle Q_{r}(t)Q_{0}(t) \right\rangle  = \left[\left\langle\mathcal{P}_r^{1} \right\rangle + \left\langle \mathcal{P}_r^{2} \right\rangle\right] + \left\langle J^{(D)}_{r}(t)Q_{0}(t)\right\rangle + \left\langle J^{(D)}_{0}(t) Q_{r}(t)\right\rangle.
\end{eqnarray}
At the steady state, we can ignore the position dependence in the averages $\left\langle\mathcal{P}_r^{1} \right\rangle$ and $\left\langle\mathcal{P}_r^{2} \right\rangle$, which leads us to write the first term in a simplified manner through the following quantity:
\begin{eqnarray}
\Gamma_{r}&=&\left\langle\mathcal{P}_r^{1} \right\rangle + \left\langle \mathcal{P}_r^{2} \right\rangle,\nonumber \\
&=&\sum_{l=\mid r \mid +1}^{\infty} \phi(l) \Bigg\{ \left(\mathcal{U}^{(l)} - \mathcal{U}^{(l+1)}\right) (l-\mid r \mid) + \sum_{g=1}^{l-1} \mathcal{V}^{(g+2)} (g-\mid r \mid)\Bigg\}\\
&=&\rho\sum_{l=\mid r \mid +1}^{\infty} \phi(l) \Bigg\{ (l-\mid r \mid)  \sum_{g=l}^{\infty}P(g) + \sum_{g=1}^{l-1} (g-\mid r \mid) P(g) \Bigg\},
\end{eqnarray}
where to arrive the last equation, which is presented as Eq.~\eqref{gamma_r} in the main text, we have used the identities shown in Eq.~\eqref{identity}. Finally using the definition of $\mathcal{C}^{Q Q}_{r}(t,t)$ and the closure approximation scheme, as shown in Eq.~\eqref{closure_approximation}, we obtain the desired time-evolution equation,
\begin{eqnarray}\label{current_current_time_evolution_eq_time_app}
\frac{d}{dt}\mathcal{C}^{Q Q}_{r}(t,t)=\Gamma_{r} - D(\rho,\gamma)\Delta_{r}\mathcal{C}^{\eta Q}_{r}(t,t) - D(\rho,\gamma)\Delta_{-r}\mathcal{C}^{\eta Q}_{-r}(t,t).
\end{eqnarray} 
By using Fourier transform in Eq.~\eqref{Inverse_Fourier_transform} in the main text, we now invert the $\mathcal{C}^{\eta Q}_{r}(t,t)$ and $\mathcal{C}^{\eta Q}_{-r}(t,t)$ in Eq.~\eqref{current_current_time_evolution_eq_time_app} in their Fourier basis, and as a result, the corresponding steady-state time-evolution equation of $\mathcal{C}^{Q Q}_{r}(t,t)$ takes the following form:
\begin{eqnarray}\label{current_current_time_evolution_eq_time_1_app}
\frac{d}{dt}\mathcal{C}^{Q Q}_{r}(t,t)=\Gamma_{r} + \frac{D(\rho,\gamma)}{L} \sum_{q}\left(2-\lambda_{qr}\right)\left(1-e^{-iq}\right)\mathcal{\tilde{C}}^{\eta Q}_{q}(t,t).
\end{eqnarray}
Eq.~\eqref{current_current_time_evolution_eq_time_1_app} is the resulting time-evolution equation, which is presented in the main text as Eq.~\eqref{Q1_Q2_evolution_compact} after integration.
\vspace{0.5 cm}
\subsection{Temporal correlation of the instantaneous bond-current}
\label{sec:app6}
Our aim in this section is to derive the expression of the steady-state temporal correlation of the instantaneous current $\mathcal{C}^{JJ}_{0}(t)$ in the long-time regime, which is presented in the main text in Eq.~\eqref{inst_current_temp_correlation_asymptotic}. Note that, for any time $t > 0$, we have already derived $\mathcal{C}^{JJ}_{0}(t)$ to obey the following equation (Eq.~\eqref{inst_current_temp_correlation_most_general} in the main text):
\begin{eqnarray}\label{inst_current_temp_correlation_most_general_app}
\mathcal{C}^{JJ}_{0}(t, 0)=-\frac{D(\rho,\gamma)}{2L}\sum_{q}f_{q} e^{-\lambda_{q}D(\rho,\gamma) t},
\end{eqnarray}
where $q=2\pi n/L$ with $n=1,2,\dots,L-1$ and the quantity $f_{q}$ at the steady state is defined as,
\begin{eqnarray}\label{f_{q}_app1}
f_{q}=\rho \sum_{l=1}^{\infty}\phi(l)\left[\sum_{g=1}^{l-1}  \lambda_{gq}P(g) +  \lambda_{lq}\sum_{g=l}^{\infty}P(g)\right],
\end{eqnarray}
here $P(g)$ is the steady-state gap-distribution function of the system. Now, if we first take the infinite system size limit, i.e. $L \rightarrow \infty$, we have the following transformations: $q \rightarrow q(x)=2\pi x$, $\lambda_{q} \rightarrow \lambda(x)$ and $f_q \rightarrow f(x)$. As a result, we can convert the summation into an integration, as shown in the following:
\begin{eqnarray}\label{inst_current_temp_correlation_most_general_int_app}
\lim_{L \rightarrow \infty} \mathcal{C}^{JJ}_{0}(t) \simeq -D(\rho,\gamma) \int_{0}^{1/2} dx f(x) e^{-\lambda(x) D(\rho,\gamma) t}
\end{eqnarray}
Interestingly, if we now take the large-time limit, i.e. $L^{2}/D \gg t \gg 1/D$ for $x >0$, the integrand in Eq.~\eqref{inst_current_temp_correlation_most_general_int_app} contributes only in the limit $x \rightarrow 0$, while it becomes vanishingly small for any other $x$ values. This effectively leads to perform a small $x$ analysis of Eq.~\eqref{inst_current_temp_correlation_most_general_int_app}. Note that, in this limit of $x \rightarrow 0$, $\lambda(x)$ is quadratic, i.e. $\lambda(x) \rightarrow 4\pi^{2}x^{2}$, $\lambda(lx) \rightarrow 4\pi^{2}l^{2}x^{2}$ and $\lambda(gx) \rightarrow 4\pi^{2}g^{2}x^{2}$. These transformations straighforwardly yield $f(x) \rightarrow 8\pi^{2}x^{2}\chi$ where $\chi$ is defined in Eq.~\eqref{chi} in the main text. Following all of the aforementioned transformations, Eq.~\eqref{inst_current_temp_correlation_most_general_int_app} in terms of a new variable $z=4\pi^{2}x^{2}Dt$ is directly reduced to the following in the limit of large t:
\begin{eqnarray}\label{inst_current_temp_correlation_most_general_app1}
\lim_{L\rightarrow \infty} \mathcal{C}^{JJ}_{0}(t) \simeq -\frac{\chi(\rho,\gamma)}{2\pi\sqrt{D(\rho,\gamma)}}t^{-3/2}\int_{0}^{\infty}dz \sqrt{z} e^{-z}. 
\end{eqnarray}
Finally, using $\int_{0}^{\infty}dz \sqrt{z} e^{-z}=\sqrt{\pi}/2$, we get the desired result presented in the main text in Eq.~\eqref{inst_current_temp_correlation_asymptotic},
\begin{eqnarray}\label{inst_current_temp_correlation_asymptotic_app}
\mathcal{C}^{JJ}_{0}(t) \simeq - \frac{\chi(\rho,\gamma)}{4\sqrt{\pi D(\rho,\gamma)}} t^{-3/2}.
\end{eqnarray}
\subsection{Derivation of the time-integrated bond-current fluctuation $\mathcal{C}^{QQ}_{0}(t,t)$}\label{sec:app7}
According to Eq.~\eqref{bond-current-fluc-0} in the main text, the steady-state bond-current fluctuation for LLG is given by,
\begin{eqnarray}\label{bond-current-fluc-0_app}
\hspace{-0.5 cm}\mathcal{C}^{QQ}_{0}(t,t) =\left[\Gamma_{0}-\frac{1}{L}\sum_{q}\left(\frac{f_{q}}{\lambda_q}\right)\right]t +\frac{1}{LD}\sum_{q}\left(\frac{f_q}{\lambda_{q}^{2}}\right)\left(1 - e^{-\lambda_{q}Dt} \right),
\end{eqnarray}
where we have defined $f_{q}$ and $\lambda_{q}$ in the main text in Eqs.~\eqref{f_{q}} and \eqref{eigen-value}, respectively. Moreover, in order to obatin $\Gamma_{0}$, we put $r=0$ in Eq.~\eqref{gamma_r} in the main text, and get
\begin{eqnarray}
\Gamma_{0}=\rho\sum_{l=1}^{\infty}\phi(l)\hspace{-0.0 cm} \Bigg\{l\sum_{g=l}^{\infty}P(g)  \hspace{-0 cm}+ \sum_{g=1}^{l-1} g P(g) \Bigg\}.
\end{eqnarray}
In order to simplify Eq.~\eqref{bond-current-fluc-0_app}, we first expand $f_q$ write
\begin{eqnarray}\label{simplification}
\sum_{q}\left(\frac{f_{q}}{\lambda_q}\right)=\rho \sum_{l=1}^{\infty}\phi(l)\left[\sum_{g=1}^{l-1}P(g)\sum_{q}\left(\frac{\lambda_{gq}}{\lambda_{q}}\right) +  \left(\sum_{q}\frac{\lambda_{lq}}{\lambda_{q}}\right)\sum_{g=l}^{\infty}P(g)\right].
\end{eqnarray}
Note that, the wave vector is given by $q=2\pi n/L$ where $n=1, 2, 3, \dots, L-1$. Therefore, the above summation over $q$, appearing at the R.H.S of the above equation, can be equivalently transformed over the integer variable $n$, which can be solved easily using MATHEMATICA to have the following simplified form:
\begin{eqnarray}
\sum_{q}\left(\frac{\lambda_{gq}}{\lambda_{q}}\right)&=&\sum_{n=1}^{L-1}\left(\frac{1-\cos\left(\frac{2\pi gn}{L}\right)}{1-\cos\left(\frac{2\pi n}{L}\right)}\right)=g(L-g),\\
\sum_{q}\left(\frac{\lambda_{lq}}{\lambda_{q}}\right)&=&\sum_{n=1}^{L-1}\left(\frac{1-\cos\left(\frac{2\pi ln}{L}\right)}{1-\cos\left(\frac{2\pi n}{L}\right)}\right)=l(L-l).
\end{eqnarray}
Applying these relations in Eq.~\eqref{simplification} drastically simplifies it and the resulting equation is given by
\begin{eqnarray}
\sum_{q}\left(\frac{f_{q}}{\lambda_q}\right) = L \Gamma_{0} - 2 \chi(\rho,\gamma).
\end{eqnarray} 
Using the above equation in Eq.~\eqref{bond-current-fluc-0_app}, we finally obtain the expression of $\mathcal{C}^{QQ}_{0}(t,t)$ used in the main text in Eq.~\eqref{bond-current-fluc-1}, which is given by 
  \begin{eqnarray}\label{bond-current-fluc-1_app}
\hspace{-0.5 cm}\mathcal{C}^{QQ}_{0}(t,t)
&=&\frac{2 \chi(\rho,\gamma)}{L}t + \frac{1}{D(\rho,\gamma)L}\sum_{q}\frac{f_{q}}{\lambda_q^{2}} \left(1 - e^{-\lambda_{q}D(\rho,\gamma)t} \right).
\end{eqnarray}
\subsection{Limiting cases of $\mathcal{C}^{QQ}_{0}(t,t)$}\label{sec:app8}
In this section, we are going to calculate $\mathcal{C}^{QQ}_{0}(t,t)$ in three distinct time-regimes, which is shown in Eq.~\eqref{cf_limit} in the main text. 
\subsubsection{Case I: $t \ll 1/D$}
It is easy to check that in this particular time regime, the second and the third terms in Eq.~\eqref{bond-current-fluc-0_app} identically cancels each other, which ultimately results the following:
\begin{eqnarray}
\mathcal{C}^{QQ}_{0}(t,t)=\Gamma_{0}t.
\end{eqnarray}
\subsubsection{Case II: $1/D \ll t \ll L^{2}/D$}
In order to calculate $\mathcal{C}^{QQ}_{0}(t,t)$ in the intermediate regime, we use the expression derived in Eq.~\eqref{bond-current-fluc-1_app} and follow the footsteps of the analysis in Appendix F. As before, for infinitely large system size, i.e. $L \rightarrow \infty$, one can convert the summation into the following integral form: 
\begin{eqnarray}\label{bond-current-fluc-1_app}
\mathcal{C}^{QQ}_{0}(t,t)
= \frac{2 \chi(\rho,\gamma)}{L}t + \frac{2}{D(\rho,\gamma)}\int_{0}^{1/2}\frac{f(x)dx}{\lambda^{2}(x)} \left(1 - e^{-\lambda(x)D(\rho,\gamma)t} \right),
\end{eqnarray}
where we have used the transformations, $q=2\pi n/L \equiv 2\pi x$, $\lambda_q \rightarrow \lambda(x)$ and $f_{q} \rightarrow f(x)$. Note that, the integrand in the above equation primarily contributes in the limit $x \rightarrow 0$ in which case, following Eqs.~\eqref{eigen-value} and \eqref{f_{q}} in the main text, we can write $\lambda(x) \simeq 4\pi^{2}x^{2}$ and $f(x) \simeq 8\pi^{2}x^{2} \chi$. Finally, using the aforementioned transformations the above equation in terms of a new variable $z=4\pi^{2}x^{2}Dt$ can be written as,
\begin{eqnarray}
\mathcal{C}^{QQ}_{0}(t,t)
= \frac{2\chi(\rho,\gamma)}{L}t + \frac{\chi(\rho,\gamma)}{\pi \sqrt{D(\rho,\gamma)}}\int_{0}^{\infty}z^{-3/2}\left(1-e^{-z}\right)dz.
\end{eqnarray}
Finally using the relation $\int_{0}^{\infty}z^{-3/2}\left(1-e^{-z}\right)dz = 2\sqrt{\pi}$ and neglecting the first term which is a subleading contributor, the leading order contribution to $\mathcal{C}^{QQ}_{0}(t,t)$ can be written as,
\begin{eqnarray}
\mathcal{C}^{QQ}_{0}(t,t)\simeq \frac{2\chi(\rho,\gamma)}{\sqrt{\pi D(\rho,\gamma)}}\sqrt{t} + \mathcal{O}(t),
\end{eqnarray}
which is presented in the main text in Eq.~\eqref{cf_limit}.
\subsubsection{Case III:$L^{2}/D \ll t$}
From Eq.~\eqref{bond-current-fluc-1_app}, it is straightforward to see that, in the limit of large $t$ such that $t \gg L^{2}/D$, the exponential term contributes nothing, whereas the second term gives a constant value and the leading order contribution comes directly from the first term,, which shows linear growth of $\mathcal{C}^{QQ}_{0}(t,t)$ with $t$, and the resulting equation becomes
\begin{eqnarray}
\mathcal{C}^{QQ}_{0}(t,t)=\frac{2\chi(\rho,\gamma)}{L}t.
\end{eqnarray}
\subsection{Scaling relation of the effective mobility $\chi(\rho,\gamma)$}
In this section, we will obtain the scaling relation for $\chi(\rho,\gamma)$ in the limit $\rho \rightarrow 0$, $\gamma \rightarrow 0$, such that the ratio $\psi=\rho/\gamma$ is finite, and calculate the corresponding scaling function $\mathcal{H}(\psi)$ shown in the main text in Eqs.~\eqref{transport_scaling} and \eqref{scaling_functions}. We begin our analysis with the expression $\chi(\rho,\gamma)$ shown in Eq.~\eqref{chi} in the main text, i.e.,
\begin{eqnarray}\label{chi-appendix}
\chi(\rho,\gamma)=\frac{\rho}{2}\sum_{l=1}^{\infty}\phi(l)\left[\sum_{g=1}^{l-1}  g^{2}P(g) +  l^{2}\sum_{g=l}^{\infty}P(g)\right].
\end{eqnarray}
Note that, the hop-length distribution $\phi(l)$ is given in Eq.~\eqref{hop-length} in the main text as,
\begin{eqnarray}\label{hop-length_app}
\phi(l) = A e^{-\gamma l},
\end{eqnarray}
where the normalization constant $A=(1-e^{-1/l_p})$. Moreover, the steady-state gap distribution function $P(g)$, which is assumed to be exponentially distrubuted for $g >0$, has the following form:
\begin{eqnarray}\label{gap_dist}
P(g) \simeq N_{*} \exp(-g/g_{*}),
\end{eqnarray}
where the prefactor $N_{*}$, as shown in the main text in Eq.~\eqref{prefactor}, is given by,
 \begin{eqnarray}\label{prefactor-appendix}
N_{*}=\left(\frac{1}{\rho}-1\right)\frac{(e^{1/g_*}-1)^{2}}{e^{1/g_*}}.
\end{eqnarray}
Now using the above expression of $P(g)$ in Eq.~\eqref{chi-appendix} and performing some algebraic manipulations, we obtain
\begin{eqnarray}\label{chi-appendix-1}
\chi(\rho,\gamma)=\frac{(1-\rho)}{2}\frac{(e^{1/g_*}-1)(e^{\gamma + 1/g_*}+1)}{(e^{\gamma + 1/g_*}-1)^{2}}.
\end{eqnarray}
Note that, the above expression of $\chi(\rho,\gamma)$ is valid for any arbitrary $\rho$ and $\gamma$. However, in the following analysis, we look at two specific cases:
\begin{itemize}
\item Case I, $\rho \rightarrow 0$ and $\gamma \rightarrow \infty:$ In this case, the typical gap-size $g_*$ is given by,
\begin{eqnarray}\label{gap_size_ssep_app1}
 g_*=\frac{1}{\rho}.
\end{eqnarray}
Now, to calculate $\chi(\rho,\gamma)$, we use Eq.~\eqref{gap_size_ssep_app1} in Eq.~\eqref{chi-appendix-1}, and with the limit $\rho \rightarrow 0$ such that $\gamma + 1/g_* \simeq \gamma \gg 1$ in consideration, we obtain
\begin{eqnarray}
\chi(\rho,\gamma)&=&\frac{(e^{\rho}-1)(1-\rho)}{2}e^{-\gamma} \\
				&\simeq & \frac{\rho(1-\rho)}{2}e^{-\gamma}=\frac{\chi^{(0)}e^{-\gamma}}{2},
\end{eqnarray}
where $\chi^{(0)}=\rho(1-\rho)$ is the particle mobility in SSEP.

\item Case II, $\rho \rightarrow 0$ and $\gamma \rightarrow 0:$ In the limit of $\rho \rightarrow 0$, $\gamma \rightarrow 0$, such that the ratio $\psi=\rho/\gamma$ is finite, we make the following transformations in Eq.~\eqref{chi-appendix-1}:
\begin{itemize}
\item the typical gap-size $g_*$ obeys the following scaling relation:
\begin{eqnarray}
g_{*} \simeq \frac{1}{\rho}\mathcal{G}(\psi),
\end{eqnarray}
where $\mathcal{G}(\psi)$ is the scaling function corresponding to $g_*$, which upon consideration of the two limiting cases is assumed to be $\mathcal{G}(\psi)=\sqrt{1+\psi}$ (see the paragraph before Eq.~\eqref{transport_scaling} in the main text),
\item all the exponential terms are approximated upto the leading order contributions, i.e.,
\begin{eqnarray}
e^{\gamma + 1/g_*}-1 &\simeq & \gamma + 1/g_* = \gamma + \rho/\mathcal{G}(\psi) = \gamma (1+\psi/\mathcal{G}(\psi)),\\
e^{1/g_*}-1 &\simeq & 1/g_* = \gamma\psi/\mathcal{G}(\psi),\\
e^{\gamma + 1/g_*}+1 &\simeq & 2.
\end{eqnarray}
\end{itemize}
Finally, by substituting the above transformation in Eq.~\eqref{chi-appendix-1}, we get the leading order contribution to $\chi(\rho,\gamma)$ in the limit $\rho \rightarrow 0$ and $\gamma \rightarrow 0$, as shown below,
\begin{eqnarray}
\chi(\rho,\gamma)& \simeq &\frac{\rho(1-\rho)}{\gamma^{2}}\frac{\mathcal{G}(\psi)}{(\psi + \mathcal{G}(\psi))^{2}}.
\end{eqnarray}
Note that, by replacing $\chi^{(0)}=\rho(1-\rho)$ in the above equation, we immediately obtain the scaling relation shown in Eq.~\eqref{transport_scaling} in the main text and the corresponding scaling function is given by,
\begin{eqnarray}
\mathcal{H}(\psi)=\frac{\mathcal{G}(\psi)}{(\psi + \mathcal{G}(\psi))^{2}}.
\end{eqnarray} 
\end{itemize}
\end{widetext}
\bibliographystyle{apsrev4-1}
\bibliography{cf}
\end{document}